\documentclass[ba]{imsart}

\makeatletter
\AtBeginDocument{%
    \def\copyrightowner@text{}
    \def\journal@name{}%
    \def\journal@issn{}
    
    \def\doi{}
}
\makeatother

\RequirePackage{etoolbox}

\volume{00}
\issue{0}
\firstpage{1}
\lastpage{1}
\usepackage{amsthm}
\usepackage[colorlinks,citecolor=blue,urlcolor=blue,filecolor=blue,backref=page]{hyperref}

\usepackage[usenames,dvipsnames,svgnames,table]{xcolor} %

\usepackage{tikz}
\usetikzlibrary{positioning}
\usetikzlibrary {shapes.geometric}
\usepackage{enumitem} %
\usepackage{rotfloat} %
\usepackage{float}
\usepackage{graphicx}
\usepackage{epsfig}
\usepackage{amssymb, amsmath}
\usepackage{verbatim}
\usepackage{natbib}
\usepackage{multirow}
\usepackage[colorinlistoftodos, textsize=scriptsize]{todonotes} %
\usepackage{subcaption} %
\usepackage[font=small]{caption}
\usepackage{caption}

\usepackage[export]{adjustbox}

\newtheorem{example}{Example}[section]

\newenvironment{remark}[1][Remark]{\begin{trivlist}
		\item[\hskip \labelsep {\bfseries #1}]}{\end{trivlist}}

\newcommand{\bea}{\begin{eqnarray*}}
	\newcommand{\eea}{\end{eqnarray*}}
\newcommand{\bean}{\begin{eqnarray}}
	\newcommand{\eean}{\end{eqnarray}}

\def\bm{\boldsymbol}

\usepackage{xspace}

\usepackage{algorithm}\numberwithin{equation}{section}
\usepackage{algpseudocode}\usepackage{arash_macros}
\algnewcommand{\IIf}[1]{\State\algorithmicif\ #1\ \algorithmicthen}

\DeclareMathOperator{\rowcompress}{row-compress}
\DeclareMathOperator{\blockcompress}{block-compress}

 \newcommand{\itrmax}{N}
\newcommand{\burnin}{N_0}
\newcommand{\Zb}{\bm{Z}}

\def\kld(#1,#2){K(#1,#2)}

\newcommand\etab{\bm \eta}

\newcommand\wb{\bm{w}}
\newcommand{\Ab}{\bm{A}}
\newcommand{\zb}{\bm{z}}
\newcommand{\vb}{\bm{v}}
\newcommand{\xb}{\bm{x}}

\newcommand\n{n}

\newcommand\pib{\bm{\pi}}

\newcommand\gamb{\bm{\gamma}}
\newcommand\kb{\bm{k}}
\newcommand\tb{\bm{t}}
\newcommand\g{g}
\newcommand\gb{\bm{\g}}
\newcommand\ub{\bm{u}}

\newcommand\fb{\bm{f}}

\newcommand\xit{\widetilde\xi}
\newcommand\Ot{\widetilde{O}}
\newcommand\htil{\widetilde{h}}
\DeclareMathOperator\gem{\mathsf{GEM}}

\DeclareMathOperator{\Beta}{\mathsf{Beta}}
\DeclareMathOperator{\unif}{\mathsf{Unif}}
\DeclareMathOperator{\bern}{\mathsf{Ber}}

\DeclareMathOperator{\sbm}{\mathsf{SBM}}

\DeclareMathOperator{\cat}{\mathsf{Cat}}

\usepackage{tcolorbox}

\newcommand\given{\,|\,}

\begin{document}

	\begin{frontmatter}
	
		\title{Hierarchical stochastic block model  for community detection in multiplex networks}
		
		\runtitle{Hierarchical stochastic block model  for multiplex networks}
		\begin{aug}
		
			\author{\fnms{Arash} \snm{Amini}\thanksref{addr2}\ead[label=e2]{aaamini@ucla.edu}}
			\and
			\author{\fnms{Marina} \snm{Paez}\thanksref{addr1}\ead[label=e1]{marina@im.ufrj.br}}
			\and
			\author{\fnms{Lizhen} \snm{Lin}\thanksref{addr2}\ead[label=e3]{lizhen.lin@nd.edu}}
			
			\runauthor{ Amini, Paez and Lin}
			
				\address[addr2]{Department of Statistics, The University of California at Los Angeles, USA
				\printead*{e2}
				}
			
			\address[addr1]{Department of Statistical Methods, Federal University of Rio de Janeiro,  Brazil
				\printead{e1} %
			}
		
				\address[addr3]{Department of Applied and Computational Mathematics and Statistics, The University of Notre Dame, USA
				\printead*{e3}
				}
			
    \end{aug}

			\begin{abstract}
			Multiplex networks have become increasingly more prevalent in many fields, and have emerged as a  powerful tool for modeling the complexity of real networks. There is a critical need for developing inference models for multiplex networks that can take into account potential dependencies across different layers, particularly when the aim is community detection.  We add to a limited literature by proposing a novel and efficient Bayesian model for community detection in multiplex networks. A key feature of our approach is the ability to model varying communities at different network layers. In contrast, many existing models assume  the same communities for all  layers. Moreover, our model automatically picks up the necessary number of communities at each layer (as validated by real data examples). This is appealing, since deciding the number of communities is a challenging aspect of community detection, and especially so in the multiplex setting, if one allows the communities to change across layers.  Borrowing ideas from hierarchical Bayesian modeling, we use a hierarchical Dirichlet  prior to model community labels  across layers, allowing dependency in their structure.  Given the community labels, a stochastic block model (SBM) is assumed for each layer. We develop an efficient slice sampler for sampling the posterior distribution of the community labels as well as the link probabilities between communities. In doing so, we address some unique challenges posed by coupling the complex likelihood of SBM with the hierarchical nature of the prior on the labels.  An extensive empirical validation is performed on simulated and real data, demonstrating the superior performance of the model over single-layer alternatives, as well as the ability to uncover interesting structures in real networks.
			\end{abstract}

\end{frontmatter}

\section{Introduction}
\label{sec:intro}

Networks, which are used to model interactions among a set of entities, %
have emerged as one of the most powerful tools for modern data analysis. The last few decades have witnessed explosions in the development of models, theory, and  algorithms for %
network analysis.  
Modern network data are often complex and heterogeneous. To model such heterogeneity, \emph{multiplex networks} \citep{BOCCALETTI20141,kievela14}, which also go by various other qualifiers (multiple-layer, multiple-slice, multi-array, multi-relational), have arisen as a useful representation of complex networks.  

 A multiplex network  typically consists of a fixed set of nodes but multiple types of edges, often  representing heterogeneous relationships as well as the dynamic nature of the edges. These networks have become increasingly prevalent in many  applications. Typical examples include various types of dynamic networks~\citep{Berlingerio-2013, 2014NatPh}  including temporal networks,  dynamic social networks~\citep{etiology} and dynamic biological networks such as protein networks~\citep{protein}.
 An example of a multiplex social network is the Twitter network where different edge information  is available such as 
 ``mention'', ``follow'' and  ``retweet''~\citep{Greene}. Another example is the co-authorship network where the authors are recorded on relationships such as ``co-publishes'', ``convenes'' and ``cites'' ~\citep{Hmimida201571}.
 Many other examples can be found in  economics~\citep{poledna2015multi},~neuroscience \citep{neuro-multi}, biology~\citep{Didier15} and so on.

Due to the ubiquity of multiplex network data, there is a critical need for developing realistic statistical models that can handle their complex structures.  There has already been growing efforts in extending static (or single-layered) statistical network models to the multiplex setting \citep{mucha2010community}. Many statistical metrics have already  been extended from basic notions such as degree and node centrality \citep{degree-centrality, structure}  to the clustering coefficient and modularity~\citep{cozzo2013clustering, modulari-multiplex}.   
 Popular latent space models~\citep{hoff:latent} have also been extended to dynamic~\citep{sarkar05a} and multiplex~\citep{gollini, salter-townshend2017} networks.
Based on such models, one can perform learning tasks such as  link probability estimation or link prediction.

 Another  task that has been extensively studied in single-layer networks is  community detection, that is, clustering nodes into groups or communities. There have already been a few works proposing models or algorithms for  community detection in multiplex and multilayer networks~\citep{mucha2010community,5992619,Kuncheva:2015,Wilson:2017:CEM,de2017community,bhattacharyya2018spectral}. The general strategy adopted is  to either transform a multiplex network into a single-layer and then apply the existing community detection algorithms, or extend a model from static networks to the multiplex setting. One of the key shortcomings of many existing methods is that communities across different layers are assumed to be the same, which is clearly restrictive and often unrealistic.  Instead, there is interest in monitoring or exploring how the communities vary across different layers. Note that there is a literature on community detection for dynamic or temporal networks mainly through developing a dynamic stochastic block model or a dynamic spectral clustering algorithm, see e.g., \cite{Liu927}, \cite{pensky2019} and \cite{matias2017}, which typically require smoothing networks over time thus assuming networks are observed over a significant number of time points. 
 
We propose a novel Bayesian community detection model, namely, the hierarchical stochastic block model  (HSBM) for multiplex networks that is fundamentally different from the existing approaches.  Specifically,  we impose a random partition prior based on the hierarchical Dirichlet process~\citep{hdp}  on  communities (or partitions) across different layers. Given these communities, a stochastic block model is assumed for each layer.  One  of the appealing features  of our  model is allowing the communities to vary across different layers of the network while being able to incorporate potential dependency across them. The hierarchical aspect of HSBM  allows  for  effective borrowing of information or strength across different layers for improved estimation. Our approach has the added advantage of being able to handle a broad class of networks by allowing the number of nodes to vary, and not necessarily imposing the nodes to be fixed across layers. In addition, HSBM inherits the desirable property of  hierarchical Dirichlet priors that allow for automatic and adaptive selection of the number of communities (for each layer) based on the data.

Our simulation study  in Section~\ref{sec:simu} confirms %
that  HSBM significantly 
outperforms a model that assumes independence of layers, especially when it comes to matching community labels across layers. In particular, the superior performance of our model over its independent single-layer counterpart is manifested by the significant improvement in the slicewise or aggregate NMI (Normalized Mutual Information)  measures.

Although  a hierarchical Dirichlet process is a natural choice for modeling dependent partition structures, extending the ideas from simple mixture models to  community structured network models is not that straightforward.  In particular,  leveraging ideas from  one of our recent work \citep{hdp-slice},  we develop a new and  efficient slice sampler for exact sampling of the posterior of HSBM for inference. We will discuss some of the technical difficulties in Remark~\ref{rem:difficulty}.

The rest of the paper is organized as follow. Section \ref{HSBM} introduces the HSBM  in details. Section \ref{sec-mcmc} is devoted to describing  a novel MCMC algorithm for inference of HSBM. Simulation study and real data analysis are presented in Sections~\ref{sec:simu} and~\ref{data}. 
The code for all the experiments is available at~\cite{hsbm-repo}. We conclude with a discussion in Section~\ref{sec:discuss}.

\section{Hierarchical stochastic block model  (HSBM)}
\label{HSBM}

Consider a multiplex network with $T$ layers (or $T$ types of edge relations) and $n_t$ nodes with labels in $[n_t] = \{1,\dots,n_t\}$ for each layer $t = 1, \ldots, T$. Denote by $\Ab_t$ the adjacency matrix of the network at layer $t$, so that an observed multiplex network consists of the collection $\Ab_t \in \{0,1\}^{n_t \times n_t}$ for $t = 1, \ldots, T$. We let $\Ab$ denote this collection and view it as a partial (or irregular) adjacency tensor. That is, $\Ab = (A_{tij}, t \in [T], i,j \in [n_t])$ and $\Ab_t = (A_{tij}, i,j \in [n_t])$ where $A_{tij} = 1$ if nodes $i$ and $j$ in layer $t$ are connected.
Our goal is to estimate the clustering or community structure of the nodes in each layer, given~$\Ab$. 

Specifically, to each node $i$, at each layer $t$, we assign a community label, encoded in a variable $\Zb = (z_{ti}) \in \nats^{T \times n_t}$, where $\nats = \{1,2,\dots,\}$ is the set of natural numbers. Let
\begin{align}
	\etab_t = (\eta_{t x y})_{x,y} \in [0,1]^{\nats \times \nats}, \quad  \eta_{txy} \equiv \eta_t(x,y),
\end{align}
 be a matrix of link probabilities between communities, indexed by $\nats^2$, for $t=1, \ldots, T$. At times, we will use the equivalent notation $\eta_t(x,y) \equiv \eta_{txy}$ to increase readability.

For example, we interpret $\eta_{t12} = \eta_t(1,2)$ as the probability of an edge being formed between a node from community $1$ and a node from community $2$ at layer~$t$. Note that we have assumed that the total number of community labels is infinite. However, for a given adjacency matrix $\Ab_t$ observed at layer $t$, the number of community labels is finite, unknown, and will be denoted by $K_t$. 

For each layer $t$, we model the distribution of the adjacency matrix $\Ab_t \in \{0,1\}^{n_t \times n_t}$ as  a stochastic block model (SBM) with membership vector $\zb_t = (z_{ti}) \in \nats^{n_t}$,  and edge probability matrix $\etab_t$, that is,
\begin{align} \label{SBM_Bernoulli}
	\Ab_t \mid \zb_t,\etab_t, \sim \sbm(\zb_t;\etab_t) 
		\iff A_{tij} \iid \bern\bigl(\,\eta_t(z_{ti}, z_{tj})\, \bigr) , \quad 1 \leq i < j \leq n_t.
\end{align}

In an SBM, the link probability between nodes $i$ and $j$ is uniquely determined by which communities these nodes belong to. In our notation, at a layer $t$, the link probability between nodes $i$ and $j$ is  given by $\eta_t(z_{ti}, z_{tj})$.
Note that  our SBM notation is slightly different from the traditional stochastic block model where the set of community labels is random. In writing $\sbm(\zb;\bm{\eta})$, we assume that $\zb$ is given and nonrandom.

If there is belief  or prior information that the community structure in one layer of the network is independent of the others, independent stochastic block models (SBM) could be assumed for the $\Ab_t$'s. This assumption is, however, too restrictive when we are dealing with networks of different kinds of relations but among the same set of nodes; or with different sets of nodes but similar types of relations. %
The other extreme is to assume that all layers in the network share the same partition---meaning that $\zb_t=\zb$ for all $t$ and the $\Ab_t$'s are conditionally independent given $\zb$.

 We believe, however, that a model that can incorporate various dependencies between these two extremes is more appropriate in many applications.  In other words, it may be desirable to allow some change in the partition structure between layers, but also impose  some kind of dependence among them. Here, we propose a model that  achieves this goal by allowing the community structures at various layers to be different but dependent, using  a hierarchical specification for the distribution of the partitions. 

\subsection{Hierarchical community prior} Before stating the prior on the labels, let us introduce a simplification, namely, that $\etab_t$ does not vary by layer. Therefore, we drop index $t$, and denote the matrix of link probabilities by $\etab$ and its elements by $\eta_{xy} \equiv \eta(x,y)$. This assumption is not necessarily restrictive since, as will become clear shortly,  
our model allows for an infinite number of communities. By imposing this restriction, we are simply stating that cluster $i$ at a certain layer corresponds to the same cluster $i$ at every other layer. We then assume independent Beta prior distributions for each element of $\etab$, given by:
\begin{align}\label{eq:eta:prior}
\etab = (\eta_{xy}) \;\iid\; \Beta(\alpha_\eta,\beta_\eta), \end{align}
where $\Beta$ stands for the usual Beta distribution.

Our key idea is to impose a dependent random partition prior on the membership labels at different layers, i.e., $\zb_t, t \in [T]$. We adopt the prior based on a hierarchical Dirichlet process (HDP). 

We assume that the reader is familiar with the HDP and its various interpretations, and in particular, the Chinese Restaurant Franchise process (CRF); see for example~\citep{hdp}. In the CRF interpretation, each layer~$t$ of the network corresponds to a restaurant, and nodes are gathered around \emph{tables}, or \emph{groups}, in each restaurant. Let $g_{ti}$ denote the group of node $i$ in restaurant (i.e., layer)~$t$. All the nodes in the same group~$g$, at the same layer~$t$, share the same dish (i.e., community) which is denoted by $k_{tg}$. Thus, given all the groups $\gb_t = (g_{ti})_i$ and \emph{group-communities} $\kb_t = (k_{tg})_g$, the label of node $i$ (at layer $t$) is uniquely determined: 
\begin{align}\label{eq:zti}
    z_{ti} \given \gb_t, \, \kb_t = k_{t,g_{ti}}.
\end{align}

\begin{figure}[t]
\begin{tabular}{cc}
     \raisebox{2mm}{Layer 1} &  
     \begin{tikzpicture}[
align=center, node distance=3mm,
community1/.style={circle, draw, very thick, 
inner sep = 0pt, minimum size=7mm},
community2/.style={rectangle, draw, fill=gray!20, very thick, inner sep = 0pt, minimum size=7mm},
community3/.style={shape aspect=1, diamond, draw, , fill=gray!45, very thick, inner sep = 0pt, minimum size=7mm},
community4/.style={
isosceles triangle, 
isosceles triangle apex angle=60,
draw, very thick, minimum size=3mm},
]
\node[community1, label=1] (1) {$a$};
\node[community1, label=2, right=of 1] (2) {$a$};
\node[community2, label=3, right=of 2] (3) {$b$};
\node[community2, label=4, right=of 3] (4) {$b$};
\node[community2, label=5, right=of 4] (5) {$b$};
\node[community3, label=6, right=of 5] (6) {$c$};
\node[community3, label=7, right=of 6] (7) {$c$};
\node[community3, label=8, right=of 7] (8) {$c$};
\end{tikzpicture}\\
    \raisebox{2mm}{Layer 2} & \begin{tikzpicture}[
align=center, node distance=3mm,
community1/.style={shape border rotate=0, regular polygon,
regular polygon sides=6,
draw, very thick, 
inner sep = 0pt, minimum size=7mm},
community3/.style={shape border rotate=0, regular polygon,
regular polygon sides=5,
draw, very thick, fill=gray!45, 
inner sep = 0pt, minimum size=7mm},
community4/.style={
shape border rotate=0,
regular polygon,
regular polygon sides=3,
draw, very thick, 
inner sep = 0pt, minimum size=7mm},
]
\node[community3, label=1] (1) {$c$};
\node[community4, label=2, right=of 1] (2) {$a$};
\node[community1, label=3, right=of 2] (3) {$a$};
\node[community3, label=4, right=of 3] (4) {$c$};
\node[community3, label=5, right=of 4] (5) {$c$};
\node[community4, label=6, right=of 5] (6) {$a$};
\end{tikzpicture}

\end{tabular}

\caption{
Toy example of the nodes of a two-layer network. The numbers are the indices of the nodes in each layer. The shape of a node represents its group and the letter symbol inside represents the community it belongs to.
}
 \label{fig:toy}
\end{figure}

\newcommand{\ourcircle}{%
    \tikz{\node[circle, draw, thick, inner sep = 0pt, minimum size=2mm] {};}%
}
\newcommand{\ourpentagon}{%
 \tikz{\node[shape border rotate=0, regular polygon, regular polygon sides=5, draw, thick, fill=gray!45, inner sep = 0pt, minimum size=2mm] {};}%
}

\newcommand{\ourhexagon}{%
 \tikz{\node[shape border rotate=0, regular polygon, regular polygon sides=6, draw, thick,  inner sep = 0pt, minimum size=2mm] {};}%
}

\newcommand{\ourtriangle}{%
 \tikz{\node[shape border rotate=0, regular polygon, regular polygon sides=3, draw, thick, inner sep = 0pt, minimum size=2mm] {};}%
}

To help the reader understand the notation of the label part of the model, we present a simple toy example:
\begin{example}
    Consider a two-layer network with $n_1 = 8$ and $ n_2 = 6$ nodes. At each layer (restaurant), the nodes are presented in a line, and are numbered from left to right, as illustrated in Figure~\ref{fig:toy}. The shape of each node represents the group (table) it belongs to and the fill, as well as its label $\in \{a,b,c\}$, represents the community (dish).
    For example, in layer 1, 
    nodes 1 and 2 are grouped together,
    represented by a circle, 
    and similarly for $\{3,4,5\}$ and $\{6,7,8\}$ represented by square and diamond, respectively.
    In this first layer, each group is associated with a different community.  
    In layer 2, we have the following groups: $\{1,4,5\}$, $\{2,6\}$ and $\{3\}$, represented by a pentagon, triangle and hexagon, respectively.
    Also, note that the groups represented by a triangle and a hexagon share the same community (dish).
    In general, groups in the same layer or in two different layers can be linked via their group-community assignment, which is encoded by $k_{t\g}$. 
    For example, the circular group in the first layer shares the same community (dish) with the triangular and hexagonal groups in the second layer. This information
    is carried through the definition of $k_{t\g}$:
    According to Figure~\ref{fig:toy},
    \[
    k_{1,\ourcircle} = k_{2,\ourtriangle} = k_{2,\ourhexagon} = a.
    \]
    This, in turn, implies that all the nodes in these groups (nodes $\{1,2\}$ in layer 1 and nodes $\{2,6\}\cup\{3\}$ in layer 2) share the same community, that is, we have 
    \[
    z_{11} = z_{12} = z_{22} = z_{26} = z_{23} = a.
    \]
    As an example of how~\eqref{eq:zti} is used to determine the node community labels, we have $g_{26} = \ourtriangle$ and $k_{2,\ourtriangle} = a$, hence $z_{26} = k_{2,g_{26}} = a$.

\end{example}

A prior on $z_{ti}$ can be obtained via 
priors on $k_{tg}$ and $g_{ti}$.
To impose dependence among layers, we assume that group-communities $\kb_t$, across layers $t$, are drawn from the same prior:
\begin{align}
k_{t\g} &\mid \pib \sim \pib, \quad \g \in \nats, 
\label{eq:ktg:prior} \\
\g_{ti} &\mid \gamb_t \sim \gamb_t, \quad i=1,\dots,n_t, \label{eq:gti:prior}
\end{align}
where $\gamb_t, t=1,\ldots, T$ are categorical  distributions with infinite categories, with the weight of each category representing the fraction of the nodes in layer $t$ that would end up in group $g$ (eventually).

To complete the prior specification, we impose
\begin{align}
\pib &\mid \gamma_0 \sim \gem(\gamma_0), \quad \pib = (\pi_k) \label{eq:pib:prior} \\
\gamb_t &\mid \alpha_0 \sim \gem(\alpha_0), \quad \gamb_t = (\gamma_{t\g}), \label{eq:gamb:prior}
\end{align}
where $\gem$ stands for  Griffiths, Engen and McCloskey \citep{picard2006combinatorial}; it is a distribution for a random measure on $\nats$ which has the well-known  stick-breaking construction \citep{sethuraman94}. Equations~\eqref{eq:pib:prior}, \eqref{eq:gamb:prior}, \eqref{eq:ktg:prior}, \eqref{eq:gti:prior} and~\eqref{eq:zti} together specify our hierarchical label prior on the node labels $(\zb_t)_{t=1}^T$.

The hierarchical label prior above, together with prior on $\etab$ in~\eqref{eq:eta:prior} and the SBM of~\eqref{SBM_Bernoulli} provide the full specification of the HSBM.

\begin{remark}
It turns out that the label prior described above is exactly the  label prior implicit in the well-known Hierarchical Dirichlet Process of~\citet{hdp}. Since this equivalence is not central to the model we present, we do not go into further details here and refer the interested reader to~\citet{hdp-slice}.
\end{remark}

\subsection{Joint distribution}
The joint density for the label part of HSBM, that is, equations~\eqref{eq:zti}--\eqref{eq:pib:prior}, can be expressed as:
\begin{align}\label{eq:joint:v11}
p(\gb,\kb,\gamb',\pib') &= \prod_{t=1}^T \Bigl[ p(\gb_t | \gamb_t) \,  p(\gamb'_t) \,p(\kb_t | \pib) \, \Bigr] p(\pib'), 
\end{align}
where $p(\gb_t | \gamb_t) = \prod_{i=1}^{n_t} \gamma_{t,\g_{ti}}$ and $p(\kb_t | \pib) = \prod_{\g=1}^\infty \pi_{k_{t\g}}$. %
The new variables 
$\gamb'_t$ and $\pib'$ are related to  $ \gamb_t$ and $ \pib$ via
the stick-breaking construction for the GEM distribution~\citep{sethuraman94,ishwaran}. The idea behind this construction is to imagine a stick with length 1, which will be successively broken into smaller pieces.  Let $F :[0,1]^\nats \to [0,1]^\nats$ be given by
 \begin{align}\label{eq:F:def}
 [F(\xb)]_1 :=  x_1, \quad [F(\xb)]_j := x_j \prod_{\ell=1}^{j-1} (1-x_\ell),
 \end{align}
 where $\xb = (x_j,j \in \nats)$. Then, 
 $[F(\xb)]_j$ is the length of the piece broken at iteration $j$  after successive fractions $x_1,x_2,\dots,x_j$ are broken off. Denote $$b_{\alpha,\beta}(x) := x^{\alpha-1} (1-x)^{\beta-1}$$ which is the density for $\Beta(\alpha,\beta)$ up to a normalization constant. 
 Both $\pib$ and $\gamb_j$ above have stick-breaking representations of this form:
 \begin{align*}
 \gamma'_{tg} &\sim b_{1,\alpha_0}(\cdot), \quad \pi'_k \sim b_{1,\gamma_0}(\cdot), \\
 \gamb_t &= F(\gamb'_t), \qquad \qquad  \pib = F(\pib'), 
 \end{align*}
 where $\gamb'_t = (\gamma'_{tg})$ and $\pib' = (\pi'_k)$.
 
 Adding the network part to the label part of the model, we obtain the full joint density of HSBM:
\begin{align}\label{eq:full:joint:v1}
\begin{split}
&p(\Ab,\etab,\gb,\kb,\gamb',\pib') = p(\Ab \mid \gb,\kb ,\etab) \cdot p(\gb,\kb,\gamb',\pib') \cdot p(\etab) \\
\qquad &= \prod_{t=1}^T \left(  
	\prod_{1 \le i < j \le n_t} L\Big(\eta(k_{t, \g_{ti}},k_{t, \g_{tj}}); A_{tij}\Big) \prod_{i=1}^{n_t} \,\gamma_{t,\g_{ti}}   \prod_{\g=1}^\infty b_{1,\alpha_0}(\gamma'_{t\g})\prod_{\g=1}^\infty \pi_{k_{t\g}} \right) \times \\
 &\qquad \prod_{k=1}^\infty  b_{1, \gamma_0}(\pi'_k) \prod_{1 \le k \le \ell < \infty} b_{\alpha_\eta,\beta_\eta}(\eta_{k \ell}),
\end{split}
\end{align}
where 
\[
L(p;a) := p^a (1-p)^{1-a}\] %
 is the Bernoulli likelihood 
  Note that we have used the alternative notation $\eta(x,y) = \eta_{xy}$ for readability. In the next section, we derive a novel MCMC algorithm for sampling the posterior distribution of our model. 
 
\section{Slice sampling for HSBM}%
\label{sec-mcmc}
We propose a slice sampler for  HSBM,  based on a slice sampling algorithm we recently developed for HDP \citep{hdp-slice}. Recall that in slice sampling from a density $f(x)$, we introduce the nonnegative variable $u$, and look at the joint density $g(x,u) = 1\{ 0 \le u \le f(x)\}$ whose marginal over $x$ is $f(x)$. Then, we perform Gibbs sampling on the joint $g$. In the end, we only keep samples of $x$ and discard those of $u$. This idea has been  employed in~\cite{Kalli11} to sample from the classical DP mixture and extended in~\cite{hdp-slice} to sample HDPs.

In order to perform the slice sampling for HSBM, we introduce (independent) variables $\ub = (u_{ti})$ and $\vb = (v_{t\g})$ so that the \emph{augmented joint density} for~\eqref{eq:joint:v11} is 
\begin{align*}
	 p(\gb,\kb,\gamb',\pib',\ub,\vb) = p(\gb, \ub \mid \gamb)\cdot p(\gamb') \cdot p(\kb, \vb \mid \pib) \cdot p(\pib'),
\end{align*}
where for example 
\[p(\gb,\ub \mid \gamb) = \prod_{t=1}^T \prod_{i=1}^{n_t} 1\{ 0 \le u_{ti} \le \gamma_{t,\g_{ti}}\}, \]
and similarly for $p(\kb,\vb \mid \pib)$. Note that marginalizing out  $\ub$ from $p(\gb,\ub \mid \gamb)$ gives back $p(\gb \mid \gamb) = \prod_t \prod_i \gamma_{t,\g_{ti}}$ as before. The full augmented joint density is now
{\small
\begin{align}\label{eq:aug:joint}
\begin{split}
	&p(\Ab,\etab,\gb,\kb,\gamb',\pib',\ub,\vb)  
	= p(\Ab \mid \gb,\kb ,\etab)  \cdot p(\gb,\kb,\gamb',\pib',\ub,\vb) \cdot p(\etab) \\
	  &= \prod_{t=1}^T \Biggl(  
	\prod_{1 \le i < j \le {n_t}} \!\!\!\!\! L\Big(\eta(k_{t, \g_{ti}},k_{t, \g_{tj}}); A_{tij}\Big) 
	\prod_{i=1}^{n_t} \,1\{u_{ti} \le \gamma_{t,\g_{ti}}\} \times \\
	&\qquad\qquad \prod_{\g=1}^\infty b_{1,\alpha_0}(\gamma'_{t\g})
	\prod_{\g=1}^\infty 1\{v_{t\g} \le \pi_{k_{t\g}}\} \Biggr) %
	\prod_{k=1}^\infty  b_{1, \gamma_0}(\pi'_k)  \!\!\!
		\prod_{1 \le k \le \ell < \infty} \!\!\!\!\! b_{\alpha_\eta,\beta_\eta}(\eta_{k \ell}),
\end{split}
\end{align}
}%
with the support understood to be restricted to $\ub \ge 0$ and $\vb \ge 0$.
We then perform block Gibbs sampling on the augmented density. 
Note that marginalizing variables $(\ub_t)$ and $(\vb_t)$ out, we get back original joint density~\eqref{eq:full:joint:v1}. The idea is to sample $(\gamb',\ub)$ jointly given the rest of the variables, and similarly for $(\pib',\vb)$.
The updates for variables $\ub, \gamb', \vb$ and $\pib'$ are similar to those in~\cite{hdp-slice}. However, the updates for the underlying latent groups $\gb$ and group-communities $\kb$ require some care due to the coupling introduced by the SBM likelihood. As can be seen from the derivation below, these updates will be quite nontrivial in the case of SBM relative to the case where the data follows a simple mixture model given the partition.

\subsection{Sampling \texorpdfstring{$(\ub,\gamb')$}{(u,gamma)}}
First, we sample $(\ub \mid \gamb', \Theta_{-\ub\gamb'})$, where $\Theta_{-\ub\gamb'}$ denotes all variables except $\ub$ and $\gamb'$. This density  factorizes and coordinate posteriors are $p(u_{ti} \mid \gamb',\Theta_{-\ub\gamb'}) \propto  1\{ u_{ti} \le \gamma_{t,\g_{ti}}\}$, that is
\begin{align*}
u_{ti} \mid \gamb',\Theta_{-\ub\gamb'} \; \sim\;  \unif(0, \gamma_{t,\g_{ti}}).
\end{align*}
Next, we sample from $(\gamb' \mid \Theta_{-\ub\gamb'})$. To do this, we first marginalize out $\ub$ in~\eqref{eq:aug:joint} which gives back~\eqref{eq:full:joint:v1}.
The corresponding posterior is, thus, proportional to~\eqref{eq:full:joint:v1} viewed only as a function of $\gamb'$. The posterior factorizes over $t$ and $\g$ and we have~\cite[Lemma~1]{hdp-slice}
\begin{align}\label{eq:gamb:p:sampling}
\gamma'_{t\g} \mid  \Theta_{-\ub\gamb'}  \sim \Beta\big(\n_\g(\gb_t)+1, \n_{>\g}(\gb_t)+\alpha_0\big),
\end{align}
where $\n_\g(\gb_t) = |\{i :\; \g_{ti} = g\} |$ and $\n_{>\g}(\gb_t)  = |\{i :\; \g_{ti} > g\} | $.

\subsection{Sampling \texorpdfstring{$(\vb,\pib')$}{(u,pi)}} 
First, we sample $(\vb \mid \pib',  \Theta_{- \vb \pib'})$ which factorizes and coordinate posteriors are $p(v_{t\g} \mid \pib',\Theta_{- \vb \pib'}) \propto  1\{v_{t\g} \le \beta_{k_{t\g}}\} $, that is
\begin{align*}
v_{t\g} \mid \pib',\Theta_{- \vb \pib'}\; \sim\;  \unif(0, \pi_{k_{t\g}}).
\end{align*}
Next, we sample from $(\pib' \mid \Theta_{- \vb \pib'})$. As in the case of $\gamb'$, we first marginalize $\vb$ which leads to the usual block Gibbs sampler updates: The posterior factorizes over $k$, and %
\begin{align}\label{eq:betab:p:sampling}
\pi'_k \mid \Theta_{- \vb \pib'} \; \sim\; \Beta\bigl(\n_k(\kb)+1, \n_{>k}(\kb)+\gamma_0\bigr),
\end{align}
where $\n_k(\kb) = |\{(t,\g) :\; k_{t\g} = k\} |$ and similarly for $n_{>k}(\kb)$.

\subsection{Sampling \texorpdfstring{$\gb$}{g}} 
This posterior factorizes over $t$ (but not over $i$). From~\eqref{eq:aug:joint}, we have 
\begin{align}\label{eq:slice:t:conditional}
\pr(\g_{ti} = \g \mid \gb_{-ti},\Theta_{-\gb}) \;\propto\; 
	\prod_{j \in [n_t] \setminus\{i\}} L\bigl(\eta(k_{t\g},k_{t, \g_{tj}}); A_{tij}\bigr)  1\{ u_{ti} \le \gamma_{t\g}\}.
\end{align}
Let $G_{ti} := \sup\{\g:\;  u_{ti} \le \gamma_{t\g} \}$. 
According to the above equation, $\g_{ti}$ given everything else will be distributed as
\begin{align*}
\g_{ti} \mid \cdots \;\sim\; \big(\, \rho_{ti}(\g) \,\big)_{\g \,\in\, [G_{ti}]}, %
\end{align*}
where, using $k_{t,\g_{tj}} = z_{tj}$ and $L(p;a) = p^a (1-p)^{1-a}$,
\begin{align*}
	\rho_{ti}(\g) &:= \prod_{j \in [n_t] \setminus\{i\}} L\big(\eta(k_{t\g}, z_{tj}); A_{tij}\big)
	= \prod_{j \in [n_t] \setminus\{i\}} \prod_{\ell} \Big[ L\big(\eta(k_{t\g}, \ell); A_{tij}\big) \Big]^{1\{ z_{tj} = \ell\}} \\
	&=  \prod_{\ell} \eta(k_{t\g}, \ell)^{\tau_{ti\ell}} [1-\eta(k_{t\g}, \ell)]^{m_{ti\ell} - \tau_{ti \ell}},
\end{align*}
with
\begin{align}\label{eq:tau:m:def}
	\tau_{t i \ell} := \sum_{j \in [n_t] \setminus\{i\}} A_{tij} 1\{z_{tj} = \ell\}, \quad m_{t i \ell} := \sum_{j \in [n_t] \setminus\{i\}}  1\{z_{tj} = \ell\}.
\end{align}

\subsection{Sampling \texorpdfstring{$\kb$}{k}} 
This posterior also factorizes over $t$ (but not over $\g$). First, note that since we are conditioning on $\gb$, we can simplify as
\begin{align*}
	\prod_{1 \le i < j \le n_t} L\big(\eta(k_{t, \g_{ti}},k_{t, \g_{tj}}); A_{tij}\big) 
	 &= \prod_{1 \le i < j \le n_t} \prod_{g,g'=1}^\infty 
	 \big[ L\big(\eta(k_{t \g},k_{t \g'}); A_{tij}\big)  \big]^{1 \{\g_{ti} = \g, \,\g_{tj} = \g' \}} \\
	 &=  \prod_{g,g'=1}^\infty  h_{t\g\g'}(k_{t\g},k_{t\g'})
\end{align*}
where
	$h_{t\g\g'}(k,\ell) := \eta(k,\ell)^{\xi_{t\g\g'}} [1-\eta(k,\ell)]^{O_{t\g\g'} - \xi_{t\g\g'}},$
\begin{align}\label{eq:xi:O:def}
	\xi_{t\g\g'} := \sum_{i,j} A_{tij} 1 \{\g_{ti} = \g, \,\g_{tj} = \g' \}, \quad 
	O_{t\g\g'} := \sum_{i,j} 1 \{\g_{ti} = \g, \,\g_{tj} = \g' \},
\end{align}
and the summations are over $1 \le i < j \le n_t$.
Then, the posterior of $\kb \mid \cdots$ factorizes over $t$, and for any fixed $t$, %
\begin{align}\label{eq:slice:k:conditional:1}
	p(\kb_{t} \mid \kb_{-t},\Theta_{-\kb}) 
	\;\propto\; \prod_{g,g'=1}^\infty 
	 h_{t\g\g'}(k_{t\g},k_{t\g'})	\prod_{\g=1}^\infty 1\{v_{t\g} \le \pi_{k_{t\g}}\}.
\end{align}
Note that $\xi_{t\g\g'}$ is not symmetric in $\g$ and $\g'$, because of the condition $i < j$ in the summation. 
Letting $\htil_{t\g\g'}(k,\ell) := h_{t\g\g'}(k,\ell) \, h_{t\g'\g}(\ell,k)$, it follows that for any fixed $t$ and $\g$:
\begin{align}\label{eq:slice:k:conditional:3}
	\pr(k_{t\g} = k \mid \kb_{-t\g},\Theta_{-\kb}) \;\propto\; 1\{v_{t\g} \le \pi_{k}\} \, 
	h_{t\g\g}(k,k)
	\prod_{\g':\;\g'\neq \g}
	\htil_{t\g\g'}(k,k_{t\g'}).
\end{align}
By the symmetry of $\eta(k,\ell)$ in its arguments, $h_{t\g\g'}(k,\ell)$ is also symmetric in $(k,\ell)$, and
\begin{align}\label{eq:htil:def}
	\begin{split}
	 \htil_{t\g\g'}(k,\ell) %
	&= \eta(k,\ell)^{\xit_{t\g\g'}} [1-\eta(k,\ell)]^{\Ot_{t\g\g'} - \xit_{t\g\g'}} 
	\end{split}
\end{align}
where $\xit_{t\g\g'}=\xi_{t\g\g'} + \xi_{t\g'\g}$ and $ \Ot_{t\g\g'}=O_{t\g\g'} + O_{t\g'\g}$. That is,
\begin{align*}%
		\xit_{t\g\g'} := \sum_{1 \le i \neq j \le n_t} A_{tij} 1 \{\g_{ti} = \g, \,\g_{tj} = \g' \}, \quad 
	\Ot_{t\g\g'} := \sum_{1 \le i \neq j \le n_t} 1 \{\g_{ti} = \g, \,\g_{tj} = \g' \}.
\end{align*}
Let us simplify the last factor in~\eqref{eq:slice:k:conditional:3} further. 
Using~\eqref{eq:htil:def}, we have
\begin{align*}
	\prod_{\g':\;\g'\neq \g} \htil_{t\g\g'}(k,k_{t\g'})  
	&= \prod_{\ell} \prod_{\g':\;\g'\neq \g} \big[\htil_{t\g\g'}(k,\ell)\big]^{1\{k_{tg'} = \ell\}} \\
	&=\prod_{\ell} \eta(k,\ell)^{\zeta_{t\g \ell}} \big[1-\eta(k,\ell) \big]^{R_{t\g \ell} - \zeta_{t\g\ell}},
\end{align*}
where
\begin{align}\label{eq:zeta:R:def}
	\zeta_{t\g \ell} := \sum_{g':\, g'\neq g} \xit_{tgg'} 1\{k_{tg'} = \ell\} , \quad R_{t\g \ell} := \sum_{g':\, g'\neq g} \Ot_{tgg'} 1\{k_{tg'} = \ell\}.
\end{align}
According to the above, $k_{t\g}$ given everything else will be  distributed as
\begin{align*}
k_{t\g} \mid \cdots \; \sim \; \big( \delta_{t\g}(k)  \big)_{k \,\in\, [K_{t\g}]},
\end{align*}
where $K_{t\g} := \sup\{k:\, v_{t\g} \le \pi_k \}$ and
\begin{align*}
	\delta_{t\g}(k) =  \eta_{kk}^{\xi_{t\g\g}} (1-\eta_{kk})^{O_{t\g\g} - \xi_{t\g\g}} \prod_{\ell} \eta_{k \ell}^{\zeta_{t\g \ell}} [1-\eta_{k \ell}]^{R_{t\g \ell} - \zeta_{t\g\ell}}.
\end{align*}

\subsection{Sampling \texorpdfstring{$\etab$}{eta}} 
Recalling that $z_{ti} = k_{t,\g_{ti}}$, the relevant part of~\eqref{eq:aug:joint} is
\begin{align*}
	\Big[\prod_{t=1}^T\prod_{1 \le i < j \le n_t} L\big(\eta(z_{ti},z_{tj}); A_{tij}\big)  \Big]
	\prod_{1 \le k \le \ell < \infty} b_{\alpha_\eta,\beta_\eta}(\eta_{k\ell}).
\end{align*}
Using the fact that $\eta(k,\ell) = \eta_{k \ell}$ is symmetric in its two arguments (that is, we treat $\eta_{k\ell}$ and $\eta_{\ell k}$ as the same variable), we have, for $k \le \ell$,
\begin{align*}
	p(\eta_{k \ell} \mid \cdots) \;\propto\;  \eta_{k\ell}^{\lambda_{k\ell}} (1-\eta_{k\ell})^{N_{ k \ell} - \lambda_{ k \ell}}  \,b_{\alpha_\eta,\beta_\eta}(\eta_{k\ell}),
\end{align*}
where
\begin{align}\label{eq:lambda:N}
	\lambda_{k\ell} = \sum_{t=1}^T\sum_{i, j} A_{tij} 1\{z_{ti} = k, z_{tj} = \ell\}, \quad 
	N_{k \ell} = \sum_{t=1}^T\sum_{i,j}  1\{z_{ti} = k, z_{tj} = \ell\},
\end{align}
and the $(i,j)$ summations are over $1 \le i < j \le n_t$ for $k=\ell$ and over $1 \le i \neq j \le n_t$ for $k \neq \ell$.
We conclude that for $k\le \ell$,
\begin{align*}
	\eta_{k\ell} \mid \cdots \;\sim \; \Beta\bigl( \,\lambda_{k\ell} + \alpha_\eta,\; N_{ k \ell} - \lambda_{k\ell} + \beta_\eta\,\bigr).
\end{align*}

\begin{rem}\label{rem:difficulty}
	Comparing with the updates in~\cite{hdp-slice}, we observe that, under HSBM, the updates for $\kb$ and $\gb$ (which is equivalent to $\tb$ in~\cite{hdp-slice})  and $\etab$ (which is equivalent to parameters of $\fb$ in~\cite{hdp-slice}) are much more complex. For the usual HDP, the data are assumed  to follow a simple mixture model where observations are independent given the labels, and each, only depends on its own label. This causes  the posterior for $\kb$, $\gb$ and the parameters of the mixture components to factorize over their coordinates. In contrast, the SBM likelihood makes each observation $A_{tij}$ dependent on two labels $z_{ti}$ and $z_{tj}$. This causes the posterior for $\kb$, $\gb$ and $\etab$ to remain coupled under HSBM. Nevertheless, the Gibbs sampling scheme above allows us to effectively sample from these coupled multivariate posteriors.
\end{rem}

\subsection{Computational speedup}
\label{sec:speedup}
Let us discuss some ideas that lead to the implementation of a fast sampler. We consider three ideas: Spare matrix computations, parallel versus sequential label updates and truncation versus slice sampling.

Many of the key computations for the slice sampler can be sped up for sparse networks, using fast sparse matrix-vector operations. Consider for example the computation of $\bm\tau_{t} = (\tau_{i\ell})$. This can be done by defining the operator ``$\rowcompress(A, \zb)$'' that takes a $A \in \reals^{n \times n}$ and a label vector $\zb \in [K]^n$ and compresses each row of matrix $A$ by summing over entries having the same label according to $\zb$, producing an $ n \times K$ matrix. Then, we will have
\begin{align}\label{eq:row:compress}
	\bm\tau_{t} = \rowcompress(\Ab_t, \zb_t).
\end{align}

Let $M_t$ be the number of nonzero elements of $\Ab_t \in \reals^{n_t \times n_t}$.
The above operator can be implemented in $O(M_t)$ operation by iterating only over the nonzero entries of $\Ab_t$ once. When $\Ab_t$ is sparse, this allows for a significant computational saving relative to the naive approach which takes $O(n_t^2)$, since in the sparse case, usually $M_t = O(n_t)$.

The operation~\eqref{eq:row:compress} is thus extremely fast if implemented \emph{in parallel}, i.e., the entire matrix $\bm\tau_t$ is computed all at once. To have a valid Gibbs sampler, however, one needs to perform row compression one row at a time in a \emph{sequential} manner, since once one element of $\zb_t$ is updated, the row compression for subsequent rows will be affected. 
The sequential row compression can be implemented with the same complexity of $O(M_t)$, but cannot benefit from parallelism, hence will potentially be slower than the the parallel version.
We refer to the two versions of the sampler as HSBM-par and HSBM-seq, respectively, based on whether the label updates are done in parallel or sequential.

To summarize, HSBM-par is doing an approximation of a valid Gibbs sampler (not an exact Gibbs sampler). HSBM-seq computes the current row compression, samples the corresponding label and then uses this new label when computing the compression for the next row. HSBM-par, however, computes the row compressions all in parallel and samples the labels all in parallel, using the current snapshot the labels. HSBM-par does not take into account the effect of sequentially sampling of labels on subsequent row-compressions.
In practice, we have found that HSBM-par performs as good as HSBM-seq and will be the default implementation used in simulations.

Similarly, computing $\bm\xit_{t} = (\xit_{tgg'})$ and $\bm\lambda = (\lambda_{k\ell})$ which are the key computations in updating $\kb$ and  $\etab$, can be done extremely fast. To see this, consider the operator ``$\blockcompress(A, \zb)$'' that returns the block compression of $A$ according to labels $\zb$, by summing all the entries of $A$ over blocks having the same row-column label pair. If $\zb \in [K]^n$, the output will be a $K \times K$ matrix and it can be computed by traversing only the nonzero entries of $A$ once. We have $\bm \xit_t = \blockcompress(\Ab_t, \gb_t)$ and $\bm\lambda = \sum_t \blockcompress(\Ab_t, \zb_t)$, ignoring the minor modifications needed depending on whether diagonal entries need to be accounted for or not. Both of these calculations can be done in $O(\sum_t M_t)$ operations.

\paragraph{Truncation sampler.}
Finally, it is possible to speed up the convergence by switching to a \emph{truncation sampler}: Instead of introducing auxiliary variables $\ub$ and $\vb$ to truncate the infinite measures automatically, one can truncate them at a fixed sufficiently large index. That is, we sample the (approximate) joint density
\begin{align}\label{eq:joint:trunc}
	\begin{split}
		p(\gb,\kb,\gamb',\pib') \approx \prod_{t=1}^T \left( \prod_{i=1}^{n_t} \gamma_{t,\g_{ti}} \prod_{\g=1}^G b_{1,\alpha_0}(\gamma'_{t\g})\prod_{\g=1}^G \pi_{k_{t\g}} \right) \prod_{k=1}^K b_{1, \gamma_0}(\pi'_k).
	\end{split}
\end{align}
where $K$ and $G$ are pre-specified large integers. The resulting Gibbs sampler is almost identical to the slice sampler with minor modifications. In particular, in the $\gb$-update, we need to replace $\rho_{ti}(g)$ with $\rho_{ti}(g) \gamma_{t,g}$ and in the $\kb$-update,  $\delta_{t\g}(k)$ with $\delta_{t\g}(k) \pi_k$. Otherwise, everything else remains the same. Empirically, we have found that the truncation sampler mixes faster than the slice sampler, hence will be our default choice in the simulations. 

\begin{figure}[t!]
	\begin{tabular}{cc}
	\includegraphics[width=0.49\textwidth]{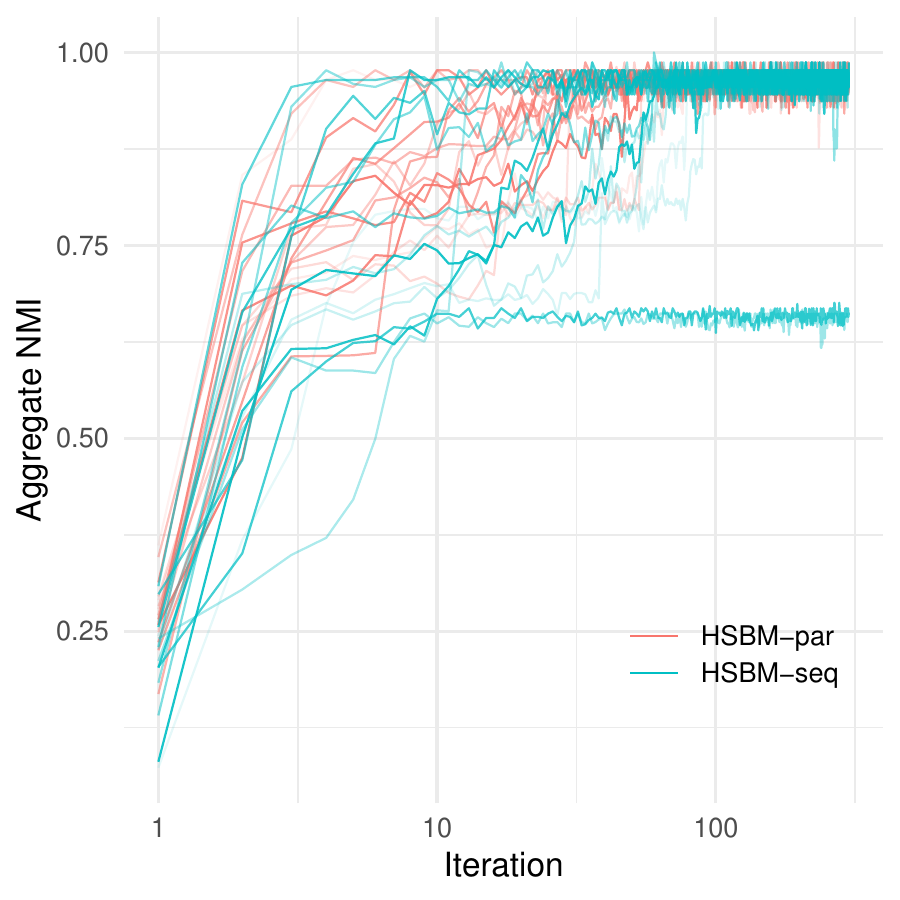} &
	\includegraphics[width=0.49\textwidth]{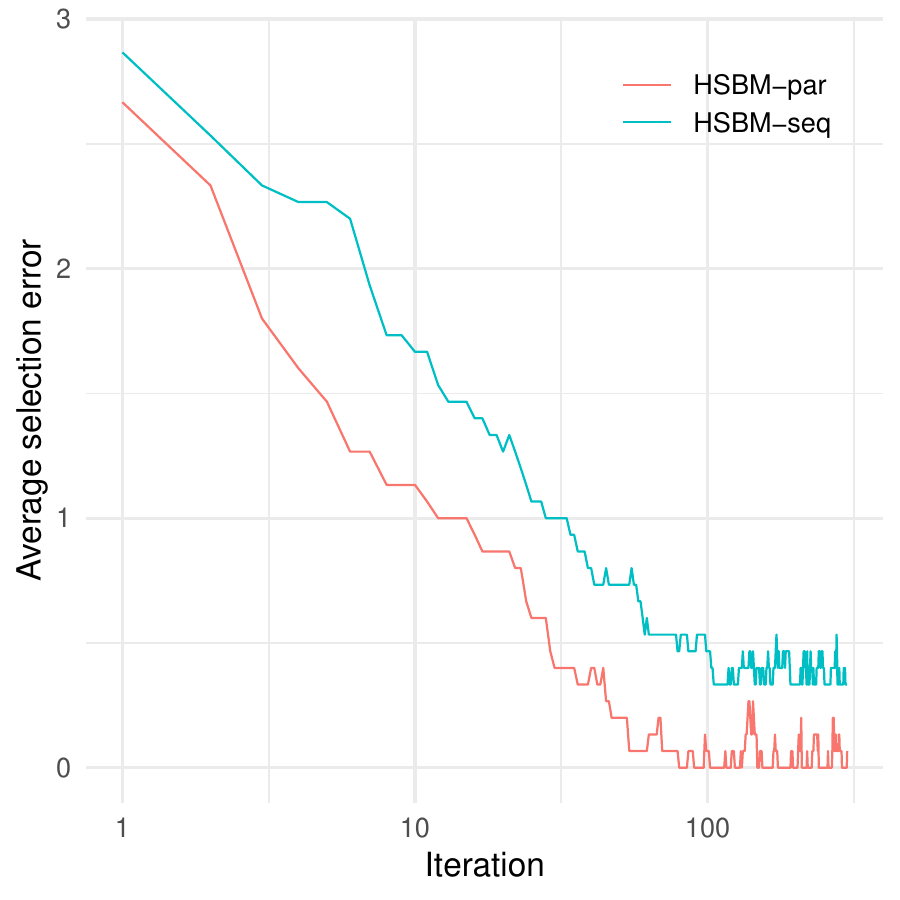} \\
	(a) & (b) 
	\end{tabular}
	\caption{The convergence speed of the sampler(s) for the personality-friendship network with $T = 5$ layers and $n_t =200$ nodes per layer. (a) Aggregate NMI relative to the true labels (b) Average error in estimating the true number of clusters.}
	\label{fig:conv}
\end{figure}

Figure~\ref{fig:conv} illustrates the convergence speed of the sampler following these implementation choices. Figure~\ref{fig:conv}(a) shows the aggregate NMI (Section~\ref{sec:nmi}) versus iteration for 15 realizations of the HSBM-par and HSBM-seq chains. The underlying network is a multilayer  personality-friendship network (Section~\ref{sec:friend:net}) with $T = 5$ layers and $n_t =200$ nodes per layer.  Figure~\ref{fig:conv}(a) is the absolute deviation between the estimated total number of communities and the truth (i.e., 3 communities), averaged over the 15 realizations. Both plots indicate fast mixing, with convergence achieved under  50 iterations for most realizations. We refer to Section~\ref{sec:simu} for more details on the simulation setup. The code for the sampler(s) is available at~\cite{hsbm-repo}.

\section{Simulation study}
\label{sec:simu}

We consider networks generated from a multilayer SBM with a single connectivity matrix, but potentially varying labels across layers. We start with an example to illustrate how such assumptions naturally arise in some applications.

\subsection{Multilayer personality-friendship network} %
\label{sec:friend:net}
Consider, as an illustration, the group of young women who run, every year, for the Miss World title. 
Each country holds a preliminary competition to choose their delegate to the main race. %
After meeting each other in the competition, some of the girls tend to bond and become friends at a social networking website. 
Suppose that we are interested in identifying groups of competitors by their personalty types: extrovert, introvert or ambivert (a balance between the former two).
 Let us also suppose that around $40\%$ of the female population %
 can be classed as extrovert, $35\%$ introvert, and $25\%$ is in the ambivert group. Assume that the distribution of personality types among the 
contestants in the %
Miss World competition is similar to that of the general population.

\begin{table}[t!]
	\centering
	\renewcommand{\arraystretch}{1.2}
	
	\begin{tabular}{r|lll}
		& Extrovert & Ambivert & Introvert \\
		\hline
		Extrovert  & 90\% & 75\% & 50\% \\
		Ambivert & 75\% & 60\% & 25\% \\
		Introvert & 50\% & 25\% & 10\% \\
	\end{tabular}
	\vskip1ex
	\caption{Hypothetical probability of friendship between personality types} %
	\label{tab_simul2}
\end{table}

It is natural to assume that extroverts have a higher chance than introverts to form bonds with other contestants. 
For this experiment, we consider the probabilities of friendship between and within groups listed in Table~\ref{tab_simul2}.
 We consider each year of the competition as a layer and the competitors represent the nodes. Alternatively, since each country is represented by a single competitor, we can consider nodes as representing different countries. The membership group of each node (i.e., its personality type) varies from layer to layer since the same country is represented by different girls in different years. However, the connection probability, i.e., the $\etab$ matrix in our notation, remains the same across layers, assuming that there is a fundamental pattern of connections among personality types rooted in human nature and invariant across years. This example thus naturally conforms to our setup of fixed $\etab$ and potentially variable $\zb_t$ over $t=1,\dots,T$.
 
\subsection{Markovian labels and random connectivity}
\label{sec:sim:settings}
 In order the systematically study the performance of HSBM and its competitors, we introduce additional degrees of freedom in generating simulated networks:
 \begin{enumerate}\itemsep=0ex
 	\item In addition to the $\etab$ introduced in Table~\ref{tab_simul2}, we also consider a random symmetric connectivity matrix with entries on and above diagonal generated i.i.d. from $\unif(0.1,0.9)$. In the experiments where the random connectivity matrix is used, all layers share the same $\etab$, however, different replications of the experiment use different randomly generated $\etab$. Note that we do not restrict $\etab$ to be assortative and the resulting random ensemble captures the full complexity possible in an SBM connectivity matrix.
 	
 	\item We generate the labels $\zb_t = (z_{ti}), t=1,\dots,T$ based on the following Markovian process: For simplicity, we assume that all layers have the same number of nodes $n$. The elements of $\zb_1$ are generated i.i.d. from a Categorical distribution with probability vector $\pib_0$, that is, $z_{1i} \sim \cat(\pib_0), i =1,\dots, n$. Subsequent labels are held fixed at previous layer value with probability $1-\tau$ or randomly generated from $\cat(\pib_0)$, that is,
 	\[
 	z_{ti} \sim (1-\tau) \delta_{z_{t-1,i}} + \tau \cat(\pib_0)
 	\] 
 	for $t=2,\dots,T$ and $i=1,\dots,n$, where $\delta_x$ is the point-mass measure at $x$. We refer to $\tau$ as the transition probability.
 	
 	\item We vary the number of layers, $T$, say from 2 to 12.
 \end{enumerate}

\subsection{Competing methods}
\label{sec:competing:methods}
We compare the performance of HSBM with a wide variety of methods. The first natural competitor is to apply DP-SBM separately to each layer.  Here, DP-SBM is a SBM with a DP prior on its labels. It can be thought of as HSBM with a single layer and with $\wb_1$ set equal to $\pib$. This model is essentially  the same as the one considered in \cite{BCD2012}.

 We also compare with various spectral approaches: SC-sliced that applies spectral clustering separately to each slice; SC-avg that applies spectral clustering to the average adjacency matrix $\bar A = \frac1T\sum_{t=1}^T A_t$; SC-ba, the debiased spectral approach of~\cite{lei2020bias}; SC-omni, the omnibus spectral approach of~\cite{levin2017central}. In addition, we consider two versions of the PisCES algorithm~\citep{Liu927} that solves an optimization problem that smooths out spectral projection matrices across time. PisCES implements the version described in~\cite{Liu927}. PisCES-sh is a modification we made where we use the same initialization to start the $k$-means clustering algorithm across different layers. Without the shared initialization, there is no guarantee of a matching between clusters obtained by $k$-means applied to the spectral representation of different layers.  

HSBM, DP-SBM, SC-sliced, SC-omni and PisCES naturally produce labels for all layers. SC-avg and SC-ba are designed to produce a single set of labels for all layers; their underlying assumption is that the labels remain the same across layers. For these two methods, we repeat the estimated label vector for all layers to obtain a multilayer label estimate.

\subsection{Measuring performance}\label{sec:nmi}

We measure the accuracy of the estimated cluster labels using
the normalized mutual information (NMI), a well-known measure of similarity between two cluster assignments. NMI takes values in $[0,1]$ where $1$ corresponds to a perfect match.
A random assignment against the truth is guaranteed to map to NMI $\approx 0$. The NMI penalizes mismatch quite aggressively. In our setting, an NMI $\approx 0.5$ corresponds to a roughly 90\% match.

In the multilayer setting, we can compute at least two NMIs: (1) The \emph{slicewise NMI} where one takes the average of NMIs computed separately for each layer, and (2) the \emph{aggregate NMI} where we consider the labels for all the layers together and compute a single NMI between competing label assignments. Here, we focus mostly on the aggregate NMI, since achieving a high aggregate NMI is more challenging, requiring consistency both within and across layers.

For the HSBM and DP-SBM the estimated labels used in the NMI calculations will be the MAP estimates, calculated as detailed below.

\paragraph{MAP estimate.} For HSBM, we compute the maximum a posteriori (MAP) label assignment by finding, for each node, the label which is most likely according to the posterior: $\argmax_k \pr( z_{ti} = k \mid \cdots)$. Associated with the MAP estimate, there is a \emph{confidence} which is the value of the posterior probability. To compute the MAP estimate we use the posterior estimate given by $\frac1{ \itrmax - \burnin} \sum_{j=\burnin+1}^\itrmax 1( \hat z^{(j)}_{ti} = k)$ where 
$\hat \zb^{(j)}= (\hat z^{(j)}_{it})$ is the label assignment at MCMC iteration $j$, $N$ is the total number of iterations and $N_0$ the length of the burn-in. Here, we ignore the potential mismatch between $\hat \zb^{(j)}$ and $\hat \zb^{(j')}$ due to the potential label-switching. In practice, all labels after MCMC convergence are coming from a single mode of the posterior as can be verified by computing the NMI between consecutive samples $\zb^{(j)}$ and $\zb^{(j+1)}$.

\subsection{Results}
\label{sec:sim:results}
\begin{figure}[t!]
	\begin{tabular}{cc}
	\includegraphics[width=0.49\textwidth]{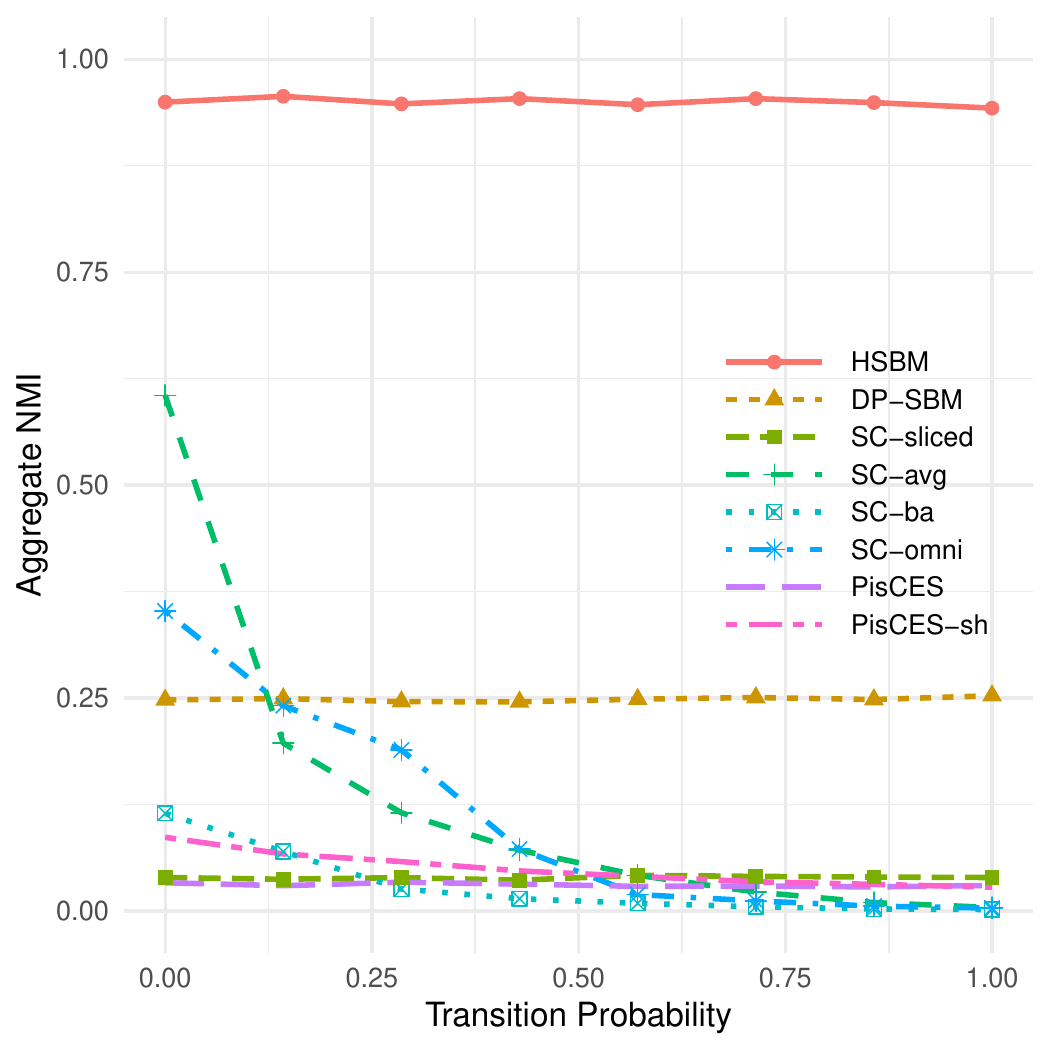} &
	\includegraphics[width=0.49\textwidth]{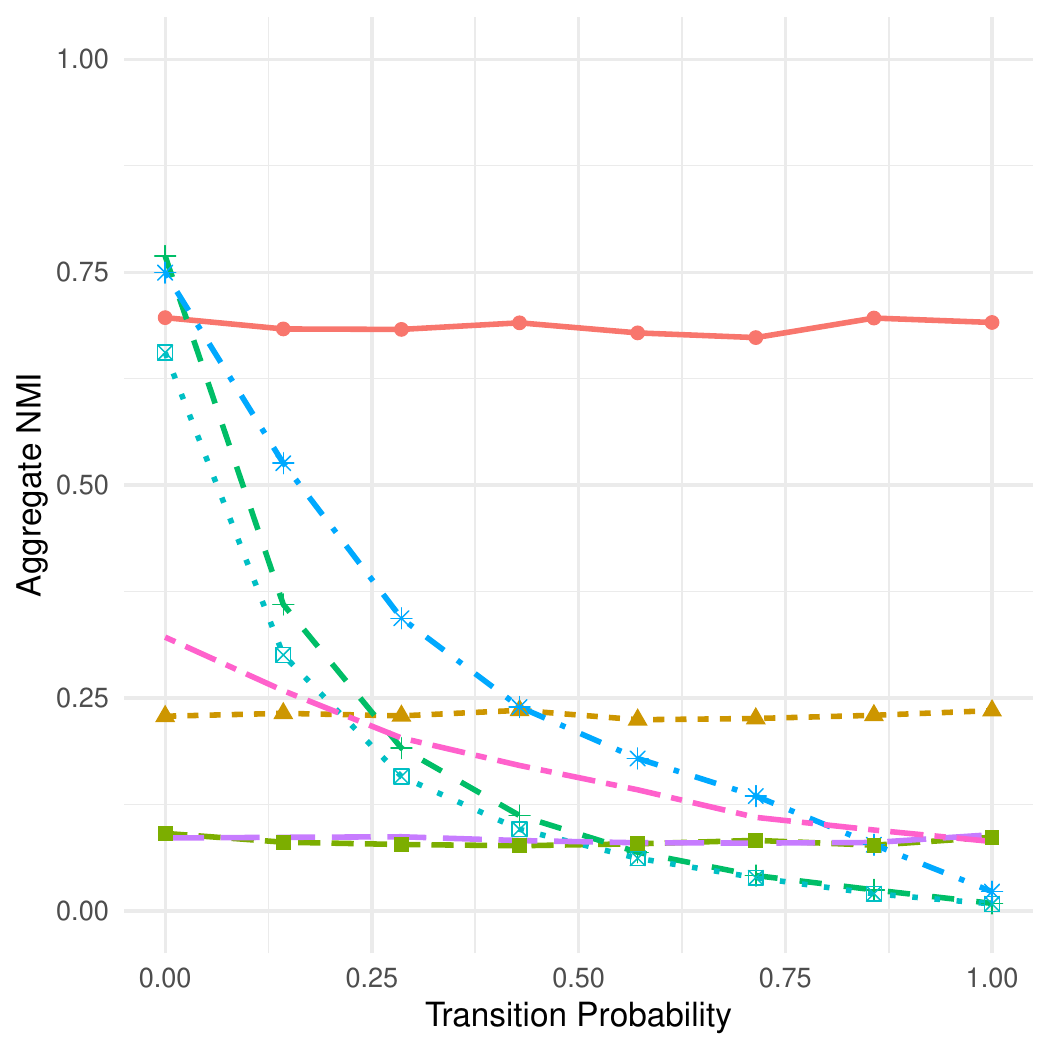} \\
	(a) & (b)
	\end{tabular}
	\caption{The performance of various methods for multilayer SBM networks with varying transition probability. (a) Personality network connectivity. (b) Random connectivity.}
	\label{fig:trans:prob} 
\end{figure}

\begin{figure}[t!]
	\begin{tabular}{cc}
		\includegraphics[width=0.48\textwidth]{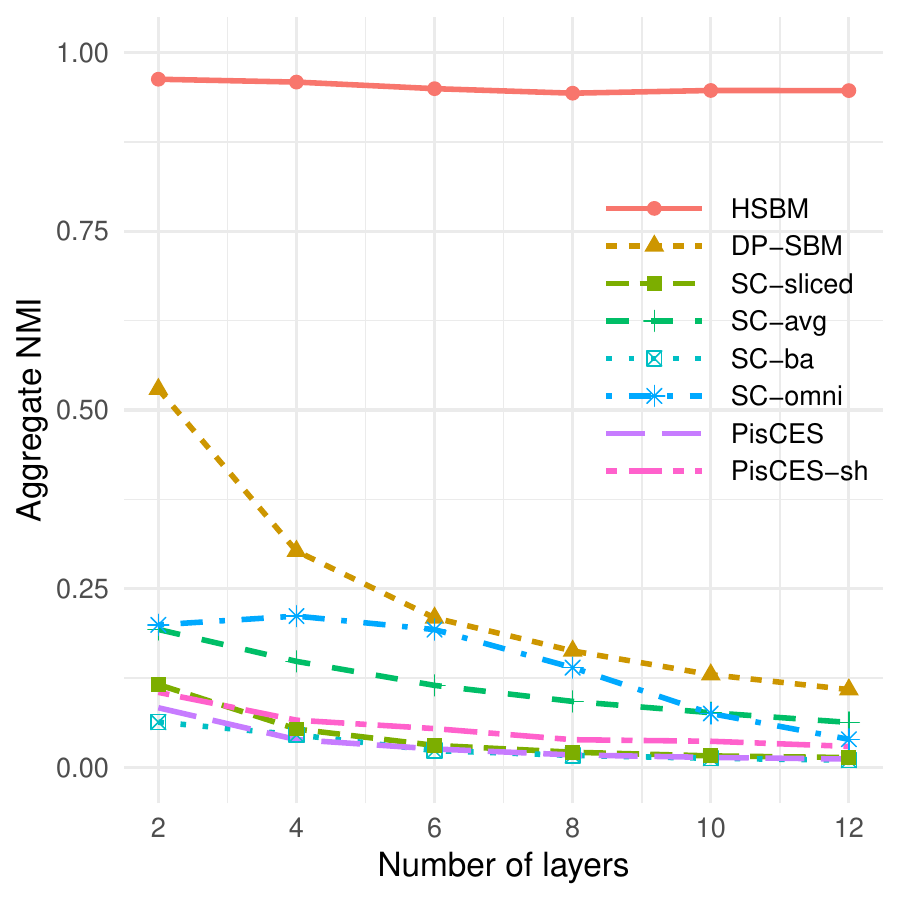} &
		\includegraphics[width=0.49\textwidth]{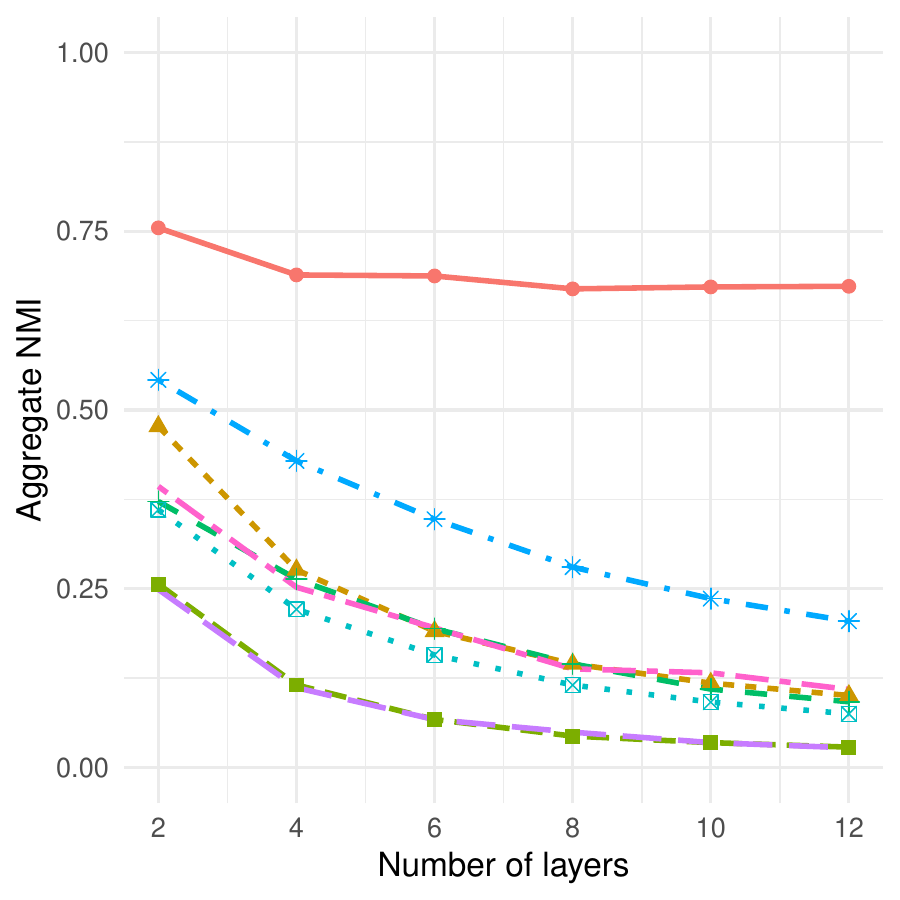} \\
		(a) & (b)
	\end{tabular}
	\caption{The performance of various methods for multilayer SBM networks with varying number of layers. (a) Personality network connectivity. (b) Random connectivity.}
	\label{fig:nlayer}
\end{figure}

We ran the HSBM and DP-SBM samplers for $N=100$ iterations with a burn-in of $N_0 = N/2$. We consider two main settings, one where we fix the number of layers at $T=5$ and change the label transition probability $\tau$, and one where we fix the transition probability at $\tau = 0.25$ and vary the number of layers $T$. In both cases, there are $n_t=200$ nodes in each layer. The results are shown in Figures~\ref{fig:trans:prob} and~\ref{fig:nlayer} respectively. In each case, the expected aggregate NMI is computed by averaging over 500 replications. For each of the two settings, we have considered both a fixed $\etab$, namely the personality-friendship connectivity of Table~\ref{tab_simul2}, and a random $\etab$ generated as discussed in Section~\ref{sec:sim:settings}. Table~\ref{tab:mean:sd} provides numerical values for the mean and the standard deviation of the aggregate and slicewise NMIs in a typical setting.
 
\begin{table}
	\renewcommand{\arraystretch}{1.2}
	\begin{tabular}{l|c|c||c|c||c|c}
		\cline{2-7}
		\multirow{2}{*}{Method} & \multicolumn{2}{c||}{Agg. NMI} & \multicolumn{2}{c||}{Slice. NMI} & \multicolumn{2}{c}{Runtime (s)}\\
		\cline{2-7}
		& mean & s.d. & mean & s.d. & mean & s.d.  \\
		\hline
		HSBM & 0.68 & 0.26 & 0.87 & 0.21 & 0.39 & 0.13\\
		DP-SBM & 0.22 & 0.07 & 0.83 & 0.17 & 1.15 & 0.34\\
		SC-omni & 0.18 & 0.09 & 0.33 & 0.11 & 0.28 & 0.14\\
		PisCES-sh & 0.14 & 0.15 & 0.56 & 0.28 & 0.18 & 0.08\\
		PisCES & 0.08 & 0.07 & 0.59 & 0.29 & 0.19 & 0.08\\
		SC-sliced & 0.08 & 0.08 & 0.56 & 0.30 & 0.10 & 0.05\\
		SC-avg & 0.07 & 0.03 & 0.11 & 0.04 & 0.05 & 0.02\\
		SC-ba & 0.06 & 0.02 & 0.12 & 0.03 & 0.15 & 0.06\\
	\end{tabular}
	\medskip
	\caption{ \small Mean and standard deviation for aggregate and slicewise NMIs as well as runtimes for a typical experiment. The setting is that of random $\etab$ with Markov labels having transition probability $\tau \approx 0.57$ and $T=5$ layers, corresponding to roughly the midpoint of the horizontal axis in Figure~\ref{fig:trans:prob}.}
	\label{tab:mean:sd}
\end{table}

The results clearly show the superior performance of HSBM relative to the competing methods. In particular, Figure~\ref{fig:trans:prob} shows that the performance of HSBM remains almost constant as one varies the transition probability ($\tau$). Most spectral methods in contrast deteriorate as $\tau$ is increased. Note that at $\tau = 1$, the labels across layers are completely independent. Nevertheless the aggregate NMI for HSBM remains high even in this case. What allows HSBM to match the clusters correctly across layers is the shared connectivity matrix $\etab$. The personality-friendship connectivity (Figure~\ref{fig:trans:prob}(a)) is especially challenging for spectral methods, mainly due to the fact the connectivity matrix is nearly rank-deficient and non-assortative. In contrast, HSBM performs almost perfectly in this case, due to the very different connectivity patterns across clusters. 

A similar qualitative behavior is observed as one varies $T$ in Figure~\ref{fig:nlayer}. The performance of most methods deteriorates as $T$ is increased while that of HSBM remains roughly the same. Interestingly, in both experiments, SC-omni is overall the most performant among the spectral methods.

\begin{remark}

Note that aggregate NMI is used as a measure of performance in most of our figures.  For a method to have a high aggregate NMI, it has to also properly match the communities across layers.  This is the main reason why  most of the SC methods fail when looking at the aggregate NMI.  For example, the SC-sliced method,  which is applied to each layer separately,   does not have any such capability.

SC-ba method is designed for the same nodes having exactly the same communities across all layers (with only the connectivity matrix changing across layers). Thus,  it fails to have high aggregate NMI in multiplex networks with varying community structures over time. 

SC-omni and PiCES allow for variable community across layers and the SC-omni is doing quite well in terms of the aggregate NMI if the transition probability is not quite high. See for example Figure 3(b).

If we only want to measure how the method performs in each layer separately, slicewise NMI is a more appropriate measure. This measure for example is reported in Table 2 and one can verify that it is quite high for SC-sliced.
On the other hand, slicewise NMI ignores the multiplex nature of the network, and that is why we focused on aggregate NMI which is a much more natural measure if one believes shared communities exist across layers.

\end{remark}

\section{Real data analysis: FAO Trade Network}
\label{data}

\begin{figure}[t]
	\begin{tabular}{cc}
		\includegraphics[width = 0.49\textwidth]{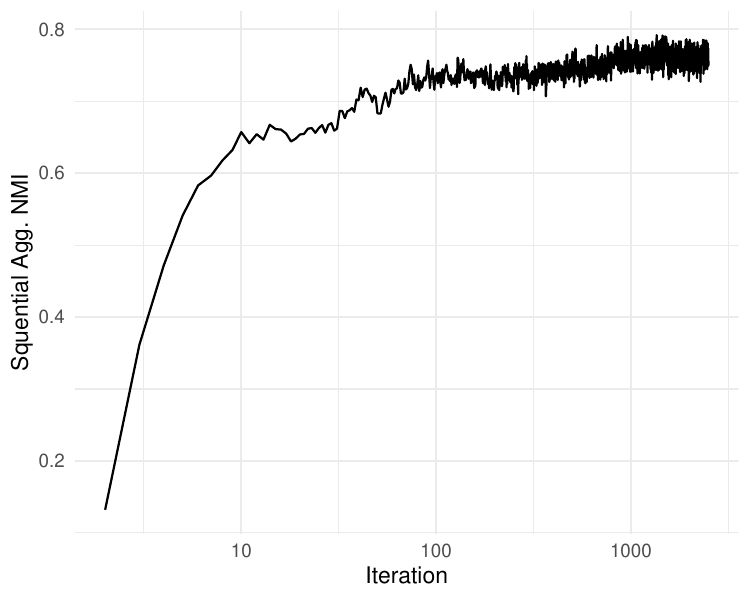} &
		\includegraphics[width = 0.49\textwidth]{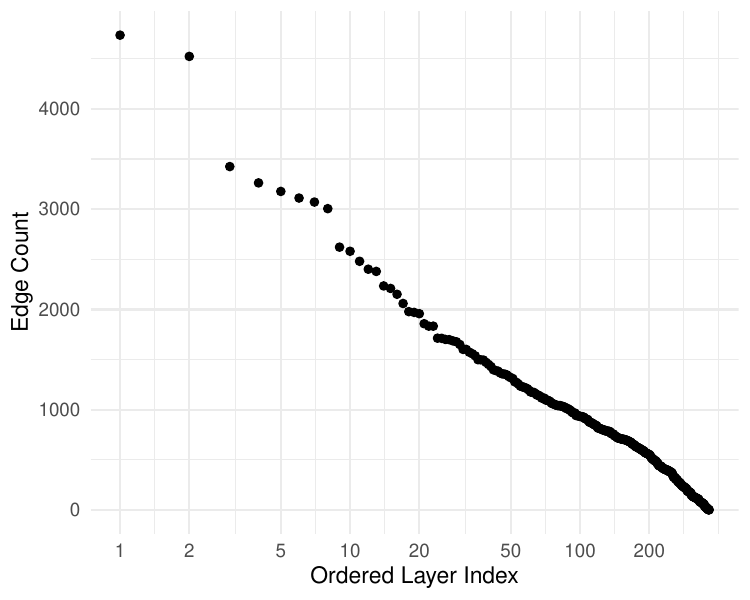} \\
		(a) & (b)
	\end{tabular}
	\caption{(a) Convergence on FAO network: Sequential Aggregate NMI versus iteration. (b)~Edge count distribution for layers of FAO network.}
	\label{fig:fao:conv:edgecount}
\end{figure}

In this section, we illustrate the performance of our model and algorithm on one real data example. 
We consider the multilayer FAO trade network provided by~\cite{de2015structural}. The data contains trade connections %
between 214 countries, treated as nodes, across 314 product categories considered as layers. We sorted the layers according to their total edge count, and chose the 20 most dense layers. Figure~\ref{fig:fao:conv:edgecount}(b) shows the distribution of the edge counts, suggesting there is a natural cut-off at about 20 layers. We then selected the nodes that had a degree greater than 20 in the \emph{sum network}, obtained by summing all the adjacency matrices. After filtering, we were left with a multilayer network on $n=172$ nodes and $20$ layers.

We then ran the HSBM sampler for 2500 iterations with a burn-in of 1250 and obtained the MAP labels. Figure~\ref{fig:fao:conv:edgecount}(a) shows the sequential NMI plot for the sampler, obtained by computing the aggregate NMI between consecutive labels across the iterations of the chain. The plot suggests that by iteration 1000 the chain is already well-mixed.

The estimated labels uncover a rich community structure.
Figure~\ref{fig:fao:net:plots} shows the community assignment across some of the layers. The nodes are laid out using a force-directed algorithm that puts highly connected nodes closer together. That the communities inferred by HSBM align with the physical proximity in these layouts is an indication of the quality of the inferred communities.

\begin{figure}[ht!]
    \centering
    \includegraphics[width=.6\textwidth]{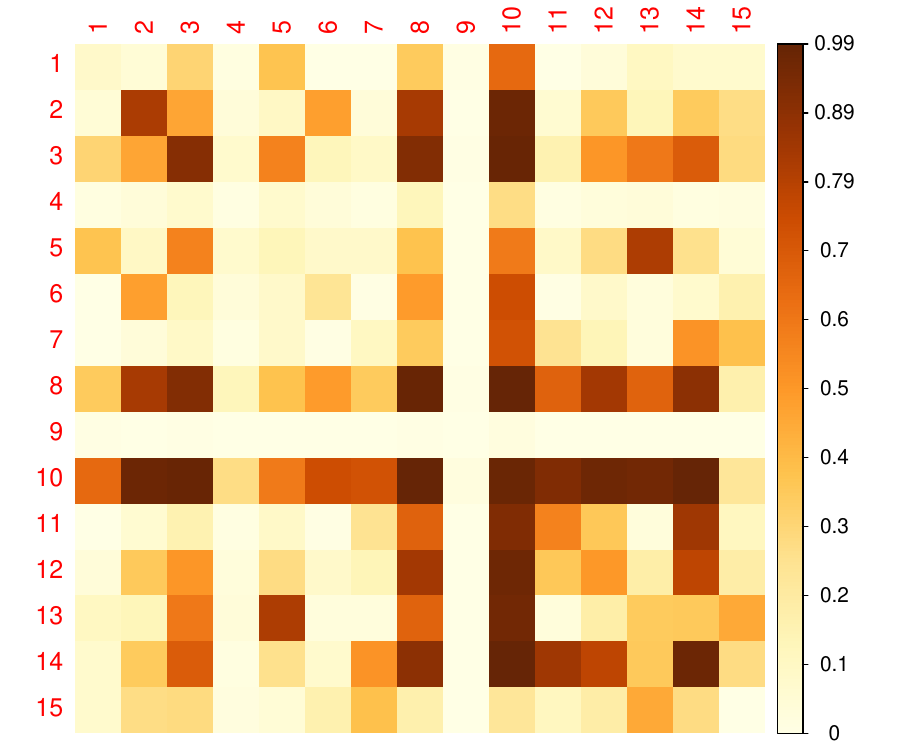}
    \caption{FAO network: The estimated connectivity matrix.}
    \label{fig:fao:conn:mat}
\end{figure}

\begin{figure}[ht!]
    \centering
    \includegraphics[width=.8\textwidth]{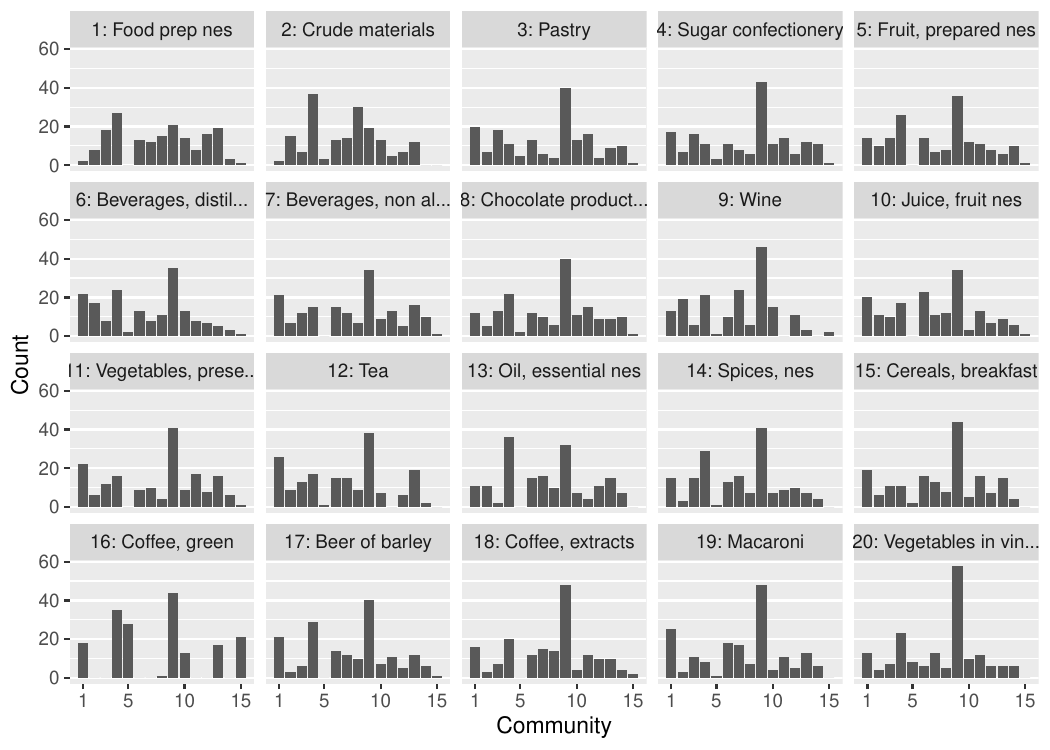}
    \caption{FAO network: The distribution of the 15 estimated communities across layers.}
    \label{fig:fao:comm_dist}
\end{figure}

There are a total of 15 estimated communities. Figure~\ref{fig:fao:conn:mat} shows the (MAP) estimated connectivity matrix $\etab$ and Figure~\ref{fig:fao:comm_dist} shows the distribution of the 15 communities across the 20 layers. Community 9 which is the most frequent across a majority of the layers corresponds to countries that sparsely trade, as evidenced by the corresponding row (and column) in $\etab$.

\begin{figure}[ht]
    \centering
    \includegraphics[width=\textwidth]{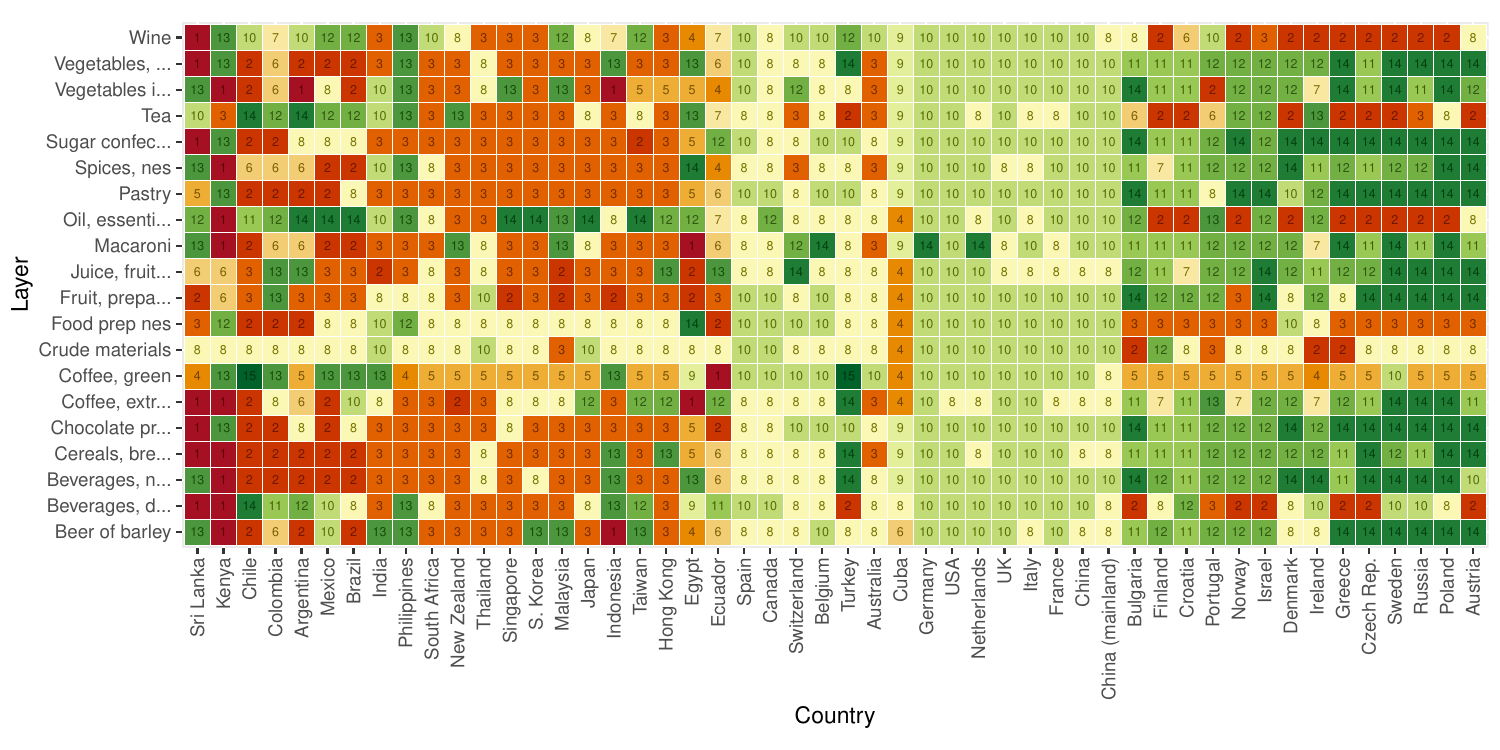}
    \caption{FAO network: Community assignments across layers for the countries that are assigned to one of the communities 8 or 10 in at least one layer.}
    \label{fig:fao:comm_table}
\end{figure}

\begin{figure}[t]
	\centering
	{\adjincludegraphics[width = 3.5cm, trim={.25\width, .3\width, .4\width, .25\width}, clip]{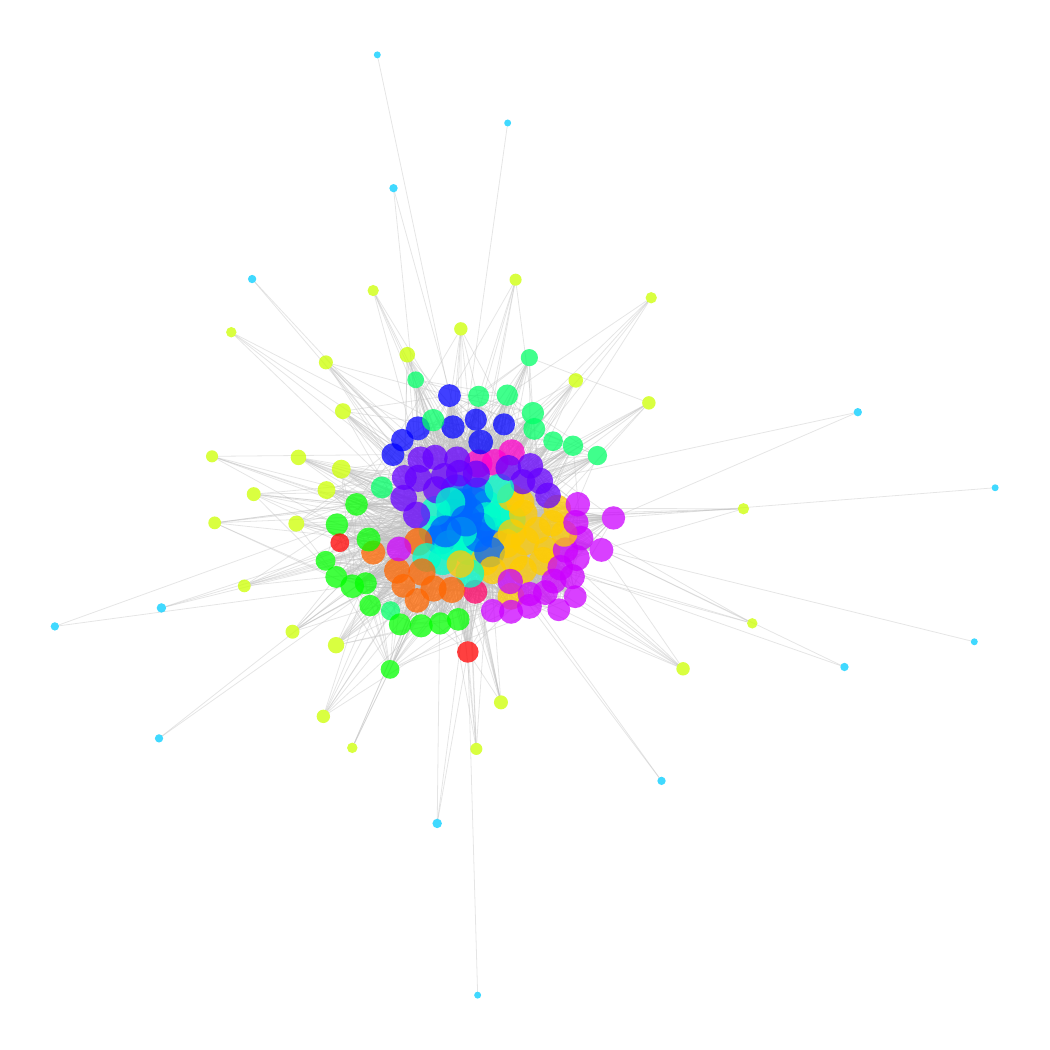}}
	{\adjincludegraphics[width = 3.5cm, trim={.4\width, .2\width, .2\width, .3\width}, clip]{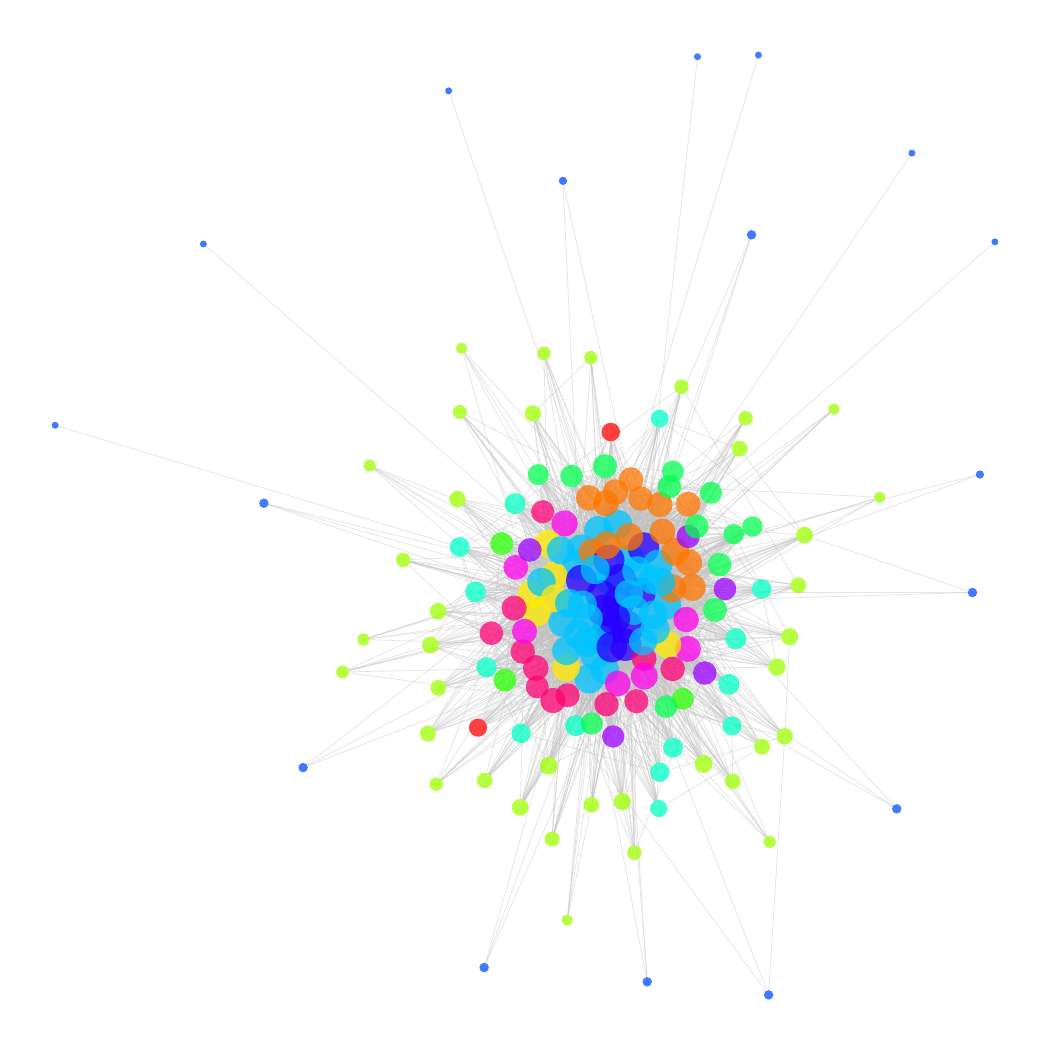}}
	{\adjincludegraphics[width = 3.5cm, trim={.35\width, .32\width, .3\width, .27\width}, clip]{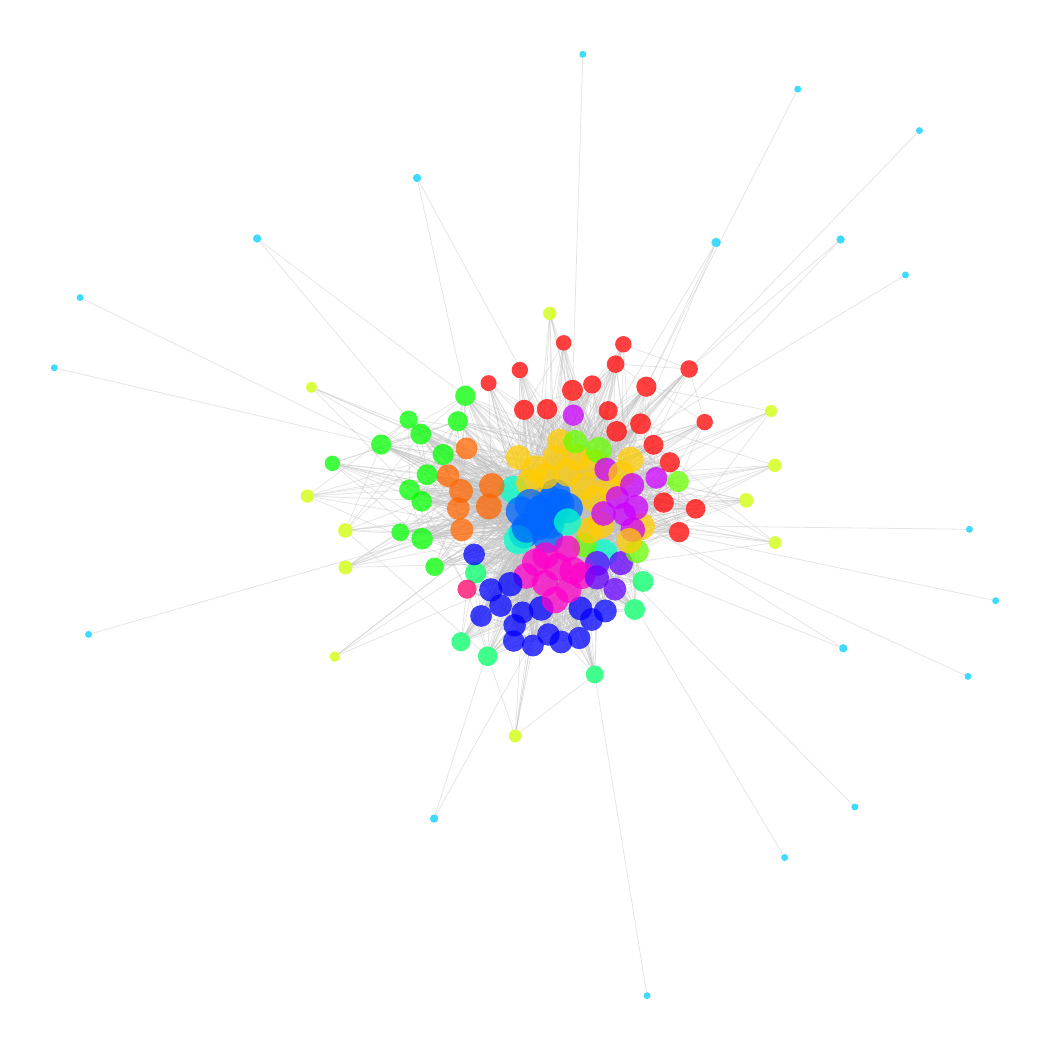}}
	{\adjincludegraphics[width = 3.5cm, trim={.3\width, .35\width, .35\width, .25\width}, clip]{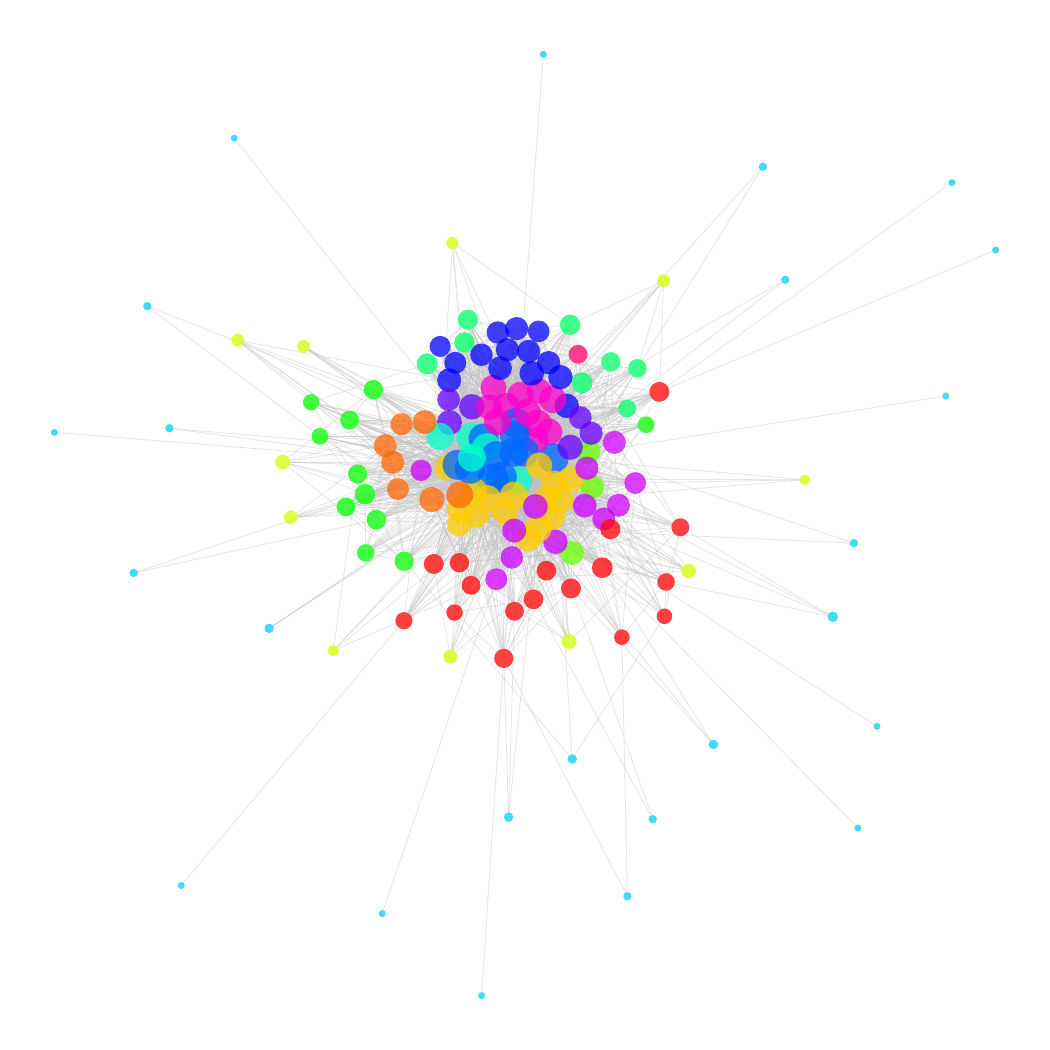}}
	{\adjincludegraphics[width = 3.5cm, trim={.3\width, .35\width, .35\width, .25\width}, clip]{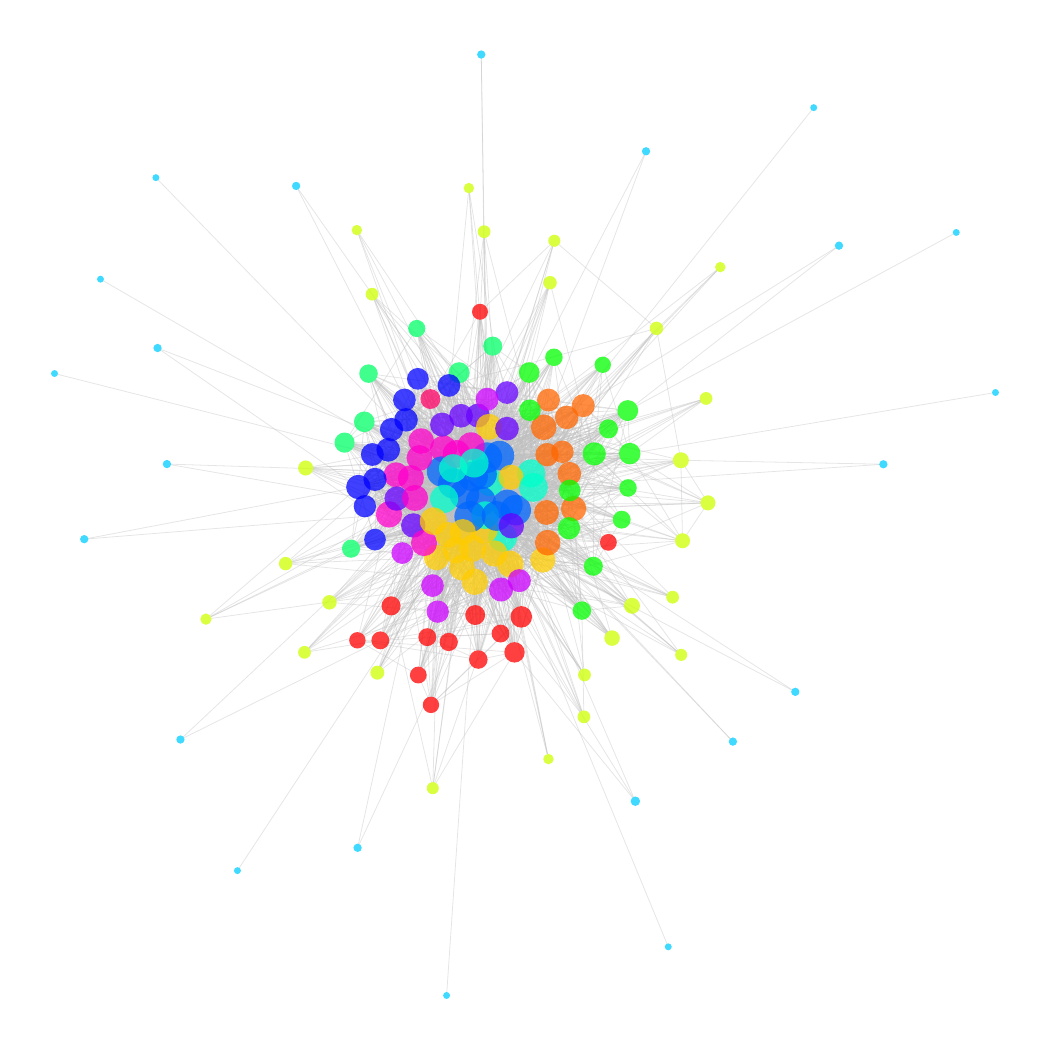}}
	{\adjincludegraphics[width = 3.5cm, trim={.35\width, .3\width, .35\width, .35\width}, clip]{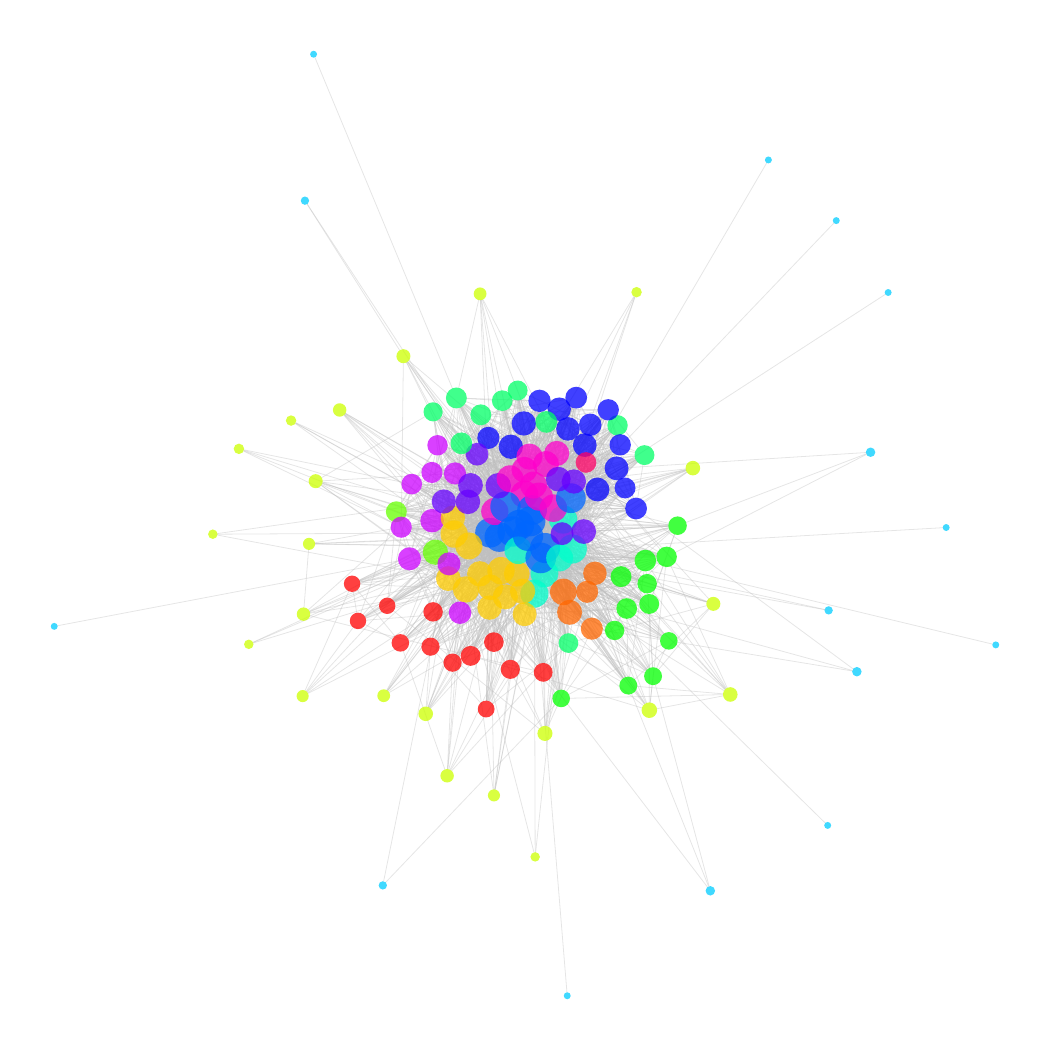}}
	\caption{FAO trade network: Examples of the community assignments. Layers shown are (from top left): Food perp. nes., Crude materials, Pastry, Suger confectionery, Fruit perp. nes., Chocolate products nes. The layouts are generated by the Fruchterman--Reingold (FR) algorithm, applied separately to each layer. FR positions the nodes according to forces exerted along the edges, resulting in spatial proximity being correlated with network connectivity. The size of a node reflects its degree. More precisely, $s_i \propto \log(d_i+3)$ where $s_i$ and $d_i$ are the size and degree of node $i$, respectively.}
	\label{fig:fao:net:plots}
\end{figure}

The connectivity matrix is not assortative--as evidenced, for example, by the near loop in Figure~\ref{fig:fao:conn:mat} among communities 10--14.  This figure also suggests that communities 8 and 10 are rather interesting. Figure~\ref{fig:fao:comm_table} shows the full community assignment for all the countries that are assigned communities 8 or 10 in at least one layer. The figure shows the complex nature of the inferred communities, with countries allowed to change communities across layers. Nevertheless, the countries that belong to group 10 across the majority of layers tend to form the tightly knit group $\{$Germany, USA, Netherlands, UK, Italy, France, China, China
 (mainland)$\}$. This is a reasonable group since these countries are known to have extensive trade relationships with each other. Interestingly, some of these countries behave like group 8 countries in some layers. Spain also has an interesting position, with equal assignments to group 10 and 8, respectively. Looking at the layers, group 8 seem to have something to do with the pattern of trade in crude materials and group 5 to the pattern of trade in green coffee. Perhaps the best way to interpret the inferred groups (or communities) is to note that they correspond to patterns of trade, hence a country can exhibit multiple of these patterns across different product. 
 
Since
under HSBM, the same country can be assigned a different label for each layer, to visualize the community assignments better, we consider the following reduction: For each label, we collect all the nodes that have that label across at least 8 out of 20 layers. Figure~\ref{fig:fao:wcloud} shows the word cloud of major groups obtained by this procedure. The size and color of a country indicate the frequency of the label across layers. For example, in Group 6, Canada has been assigned label 6 for 13 out of 20 layers, more than any other country in that group. Switzerland, Australia and Belgium were all assigned label 6 in 11 layers. In addition to the groups shown in the figure, we have three singleton groups: Chile, UAE and ``unspecified''.

\def\wcscale{0.24}
\begin{figure}[t]
		\includegraphics[width = \wcscale\textwidth]{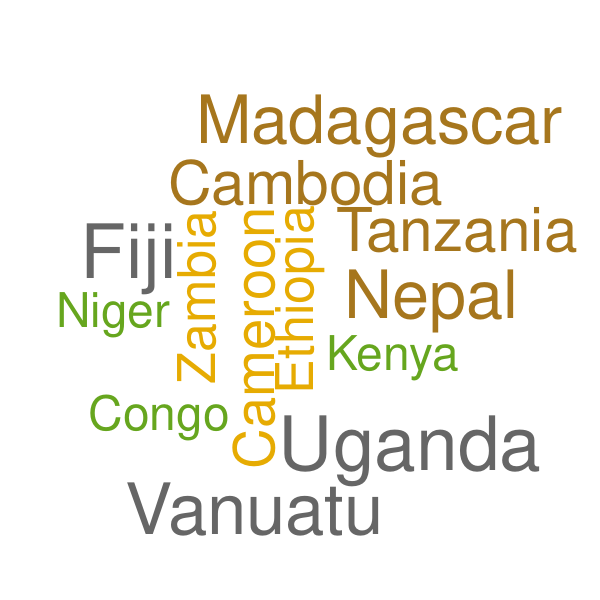} 
		\includegraphics[width = \wcscale\textwidth]{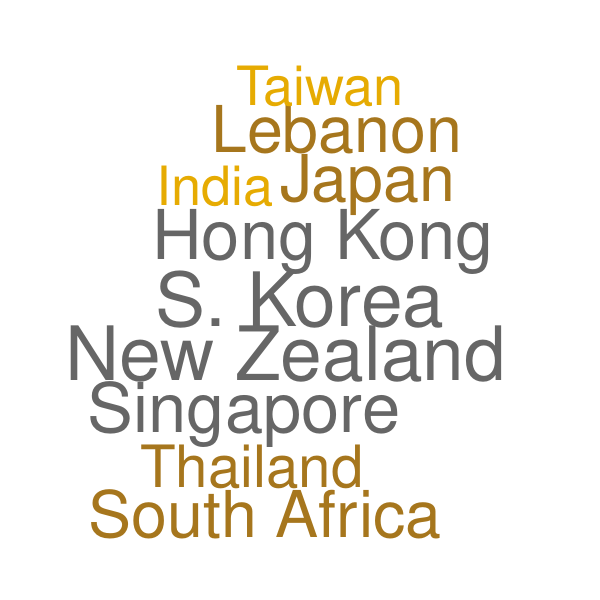} 
		\includegraphics[width = \wcscale\textwidth]{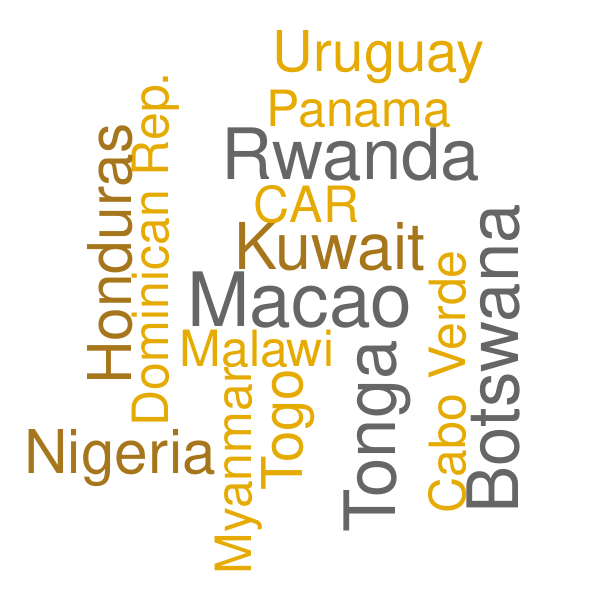}
		\includegraphics[width = \wcscale\textwidth]{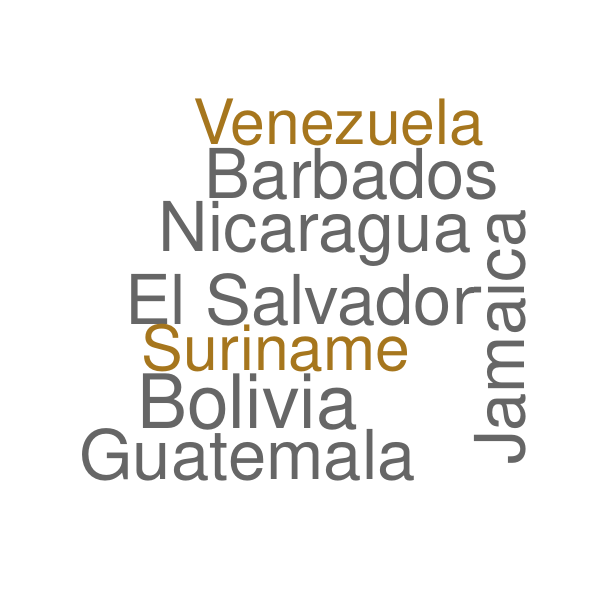}
		\includegraphics[width = \wcscale\textwidth]{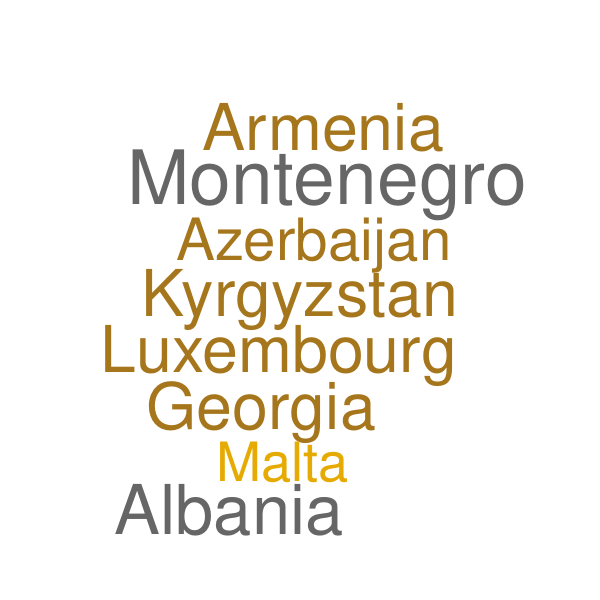}
		\includegraphics[width = \wcscale\textwidth]{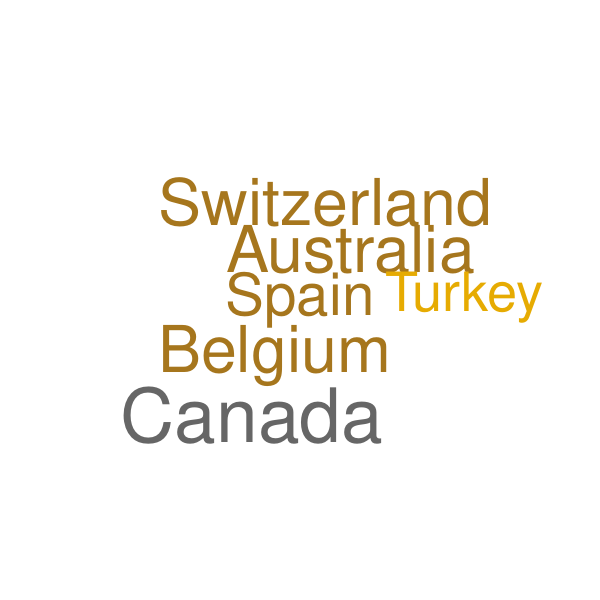}
		\includegraphics[width = \wcscale\textwidth]{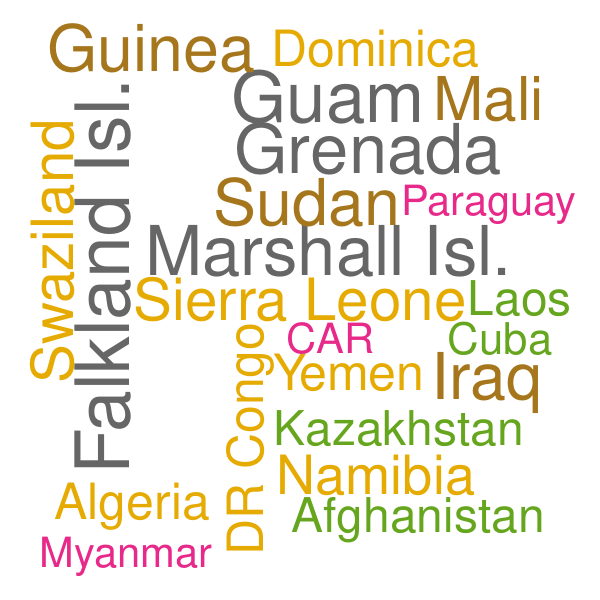}
		\includegraphics[width = \wcscale\textwidth]{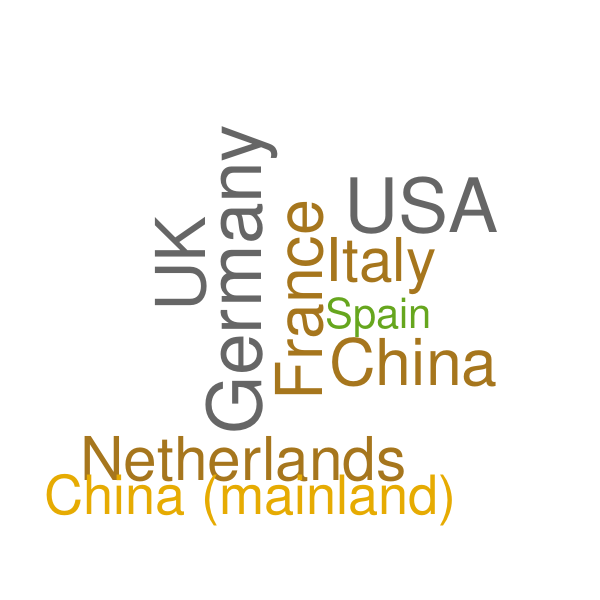}
		\includegraphics[width = \wcscale\textwidth]{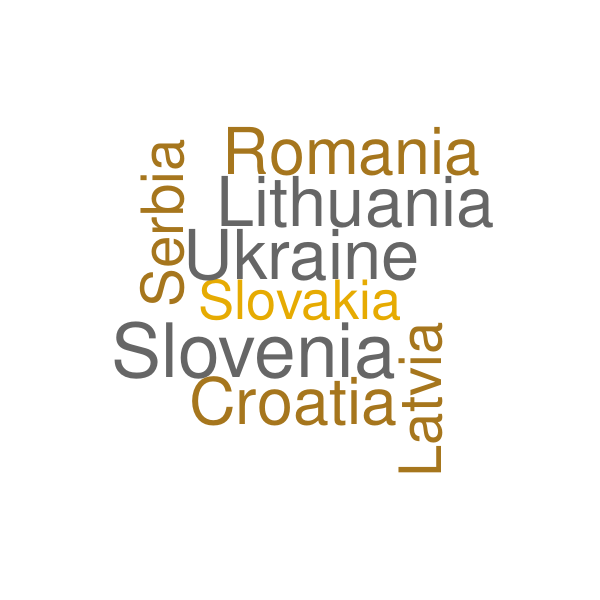}
		\includegraphics[width = \wcscale\textwidth]{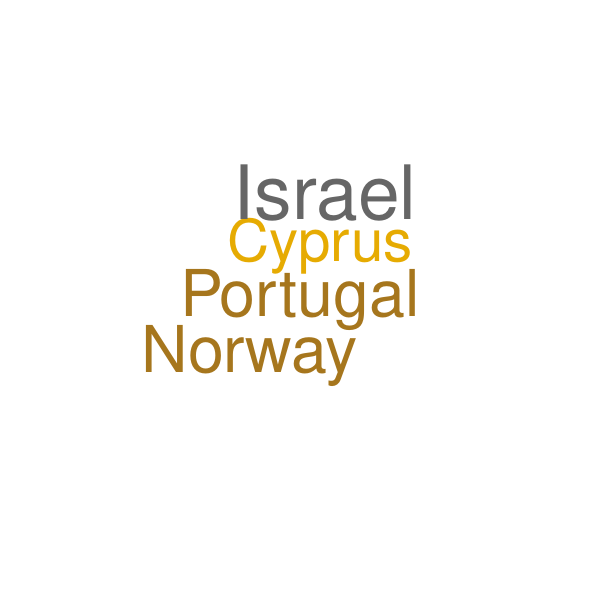}
		\includegraphics[width = \wcscale\textwidth]{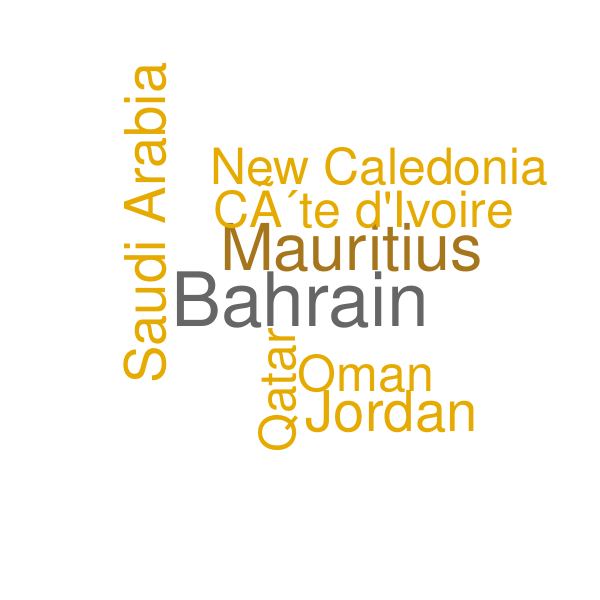}
		\includegraphics[width = \wcscale\textwidth]{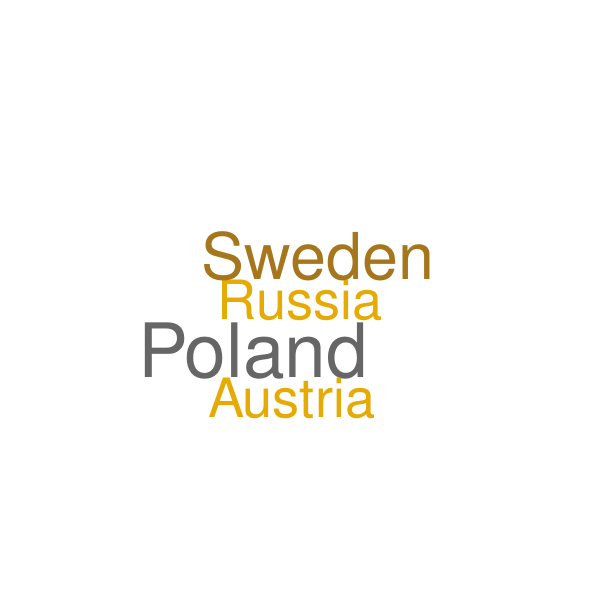}
		\caption{FAO trade network: Word cloud of groups based on frequency of assignment across layers.}
		\label{fig:fao:wcloud}
\end{figure}

\begin{figure}[t]
	\centering
	\begin{tabular}{cc}
		\includegraphics[width = 0.49\textwidth]{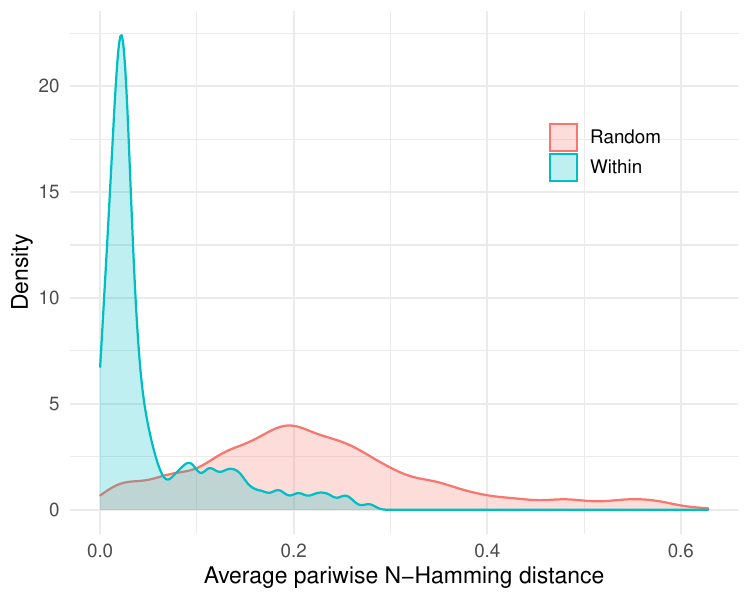} &
		\includegraphics[width = 0.49\textwidth]{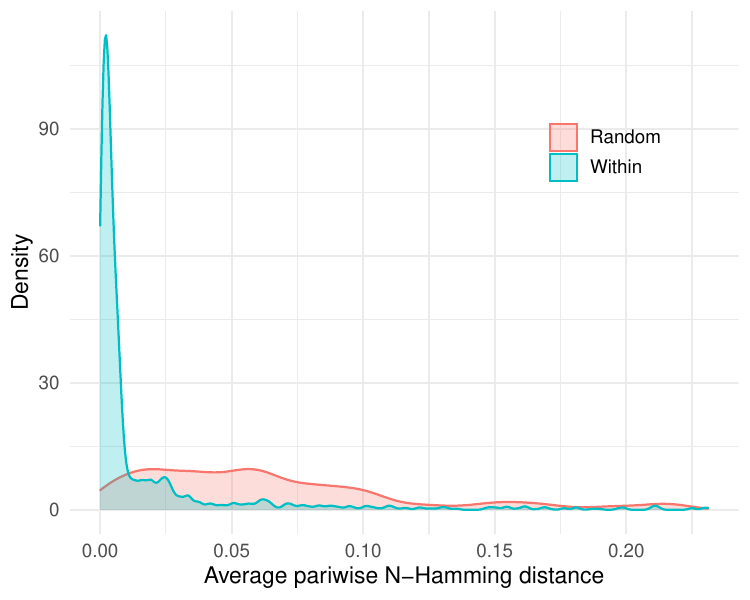} \\
		(a) & (b)
	\end{tabular}
	\caption{FAO trade network:  Average Normalized Hamming (ANH) distance between pairs of nodes. The two distributions correspond to random selection of pairs, versus (random) selection within estimated groups. (a) Calculated based on the in-sample layers. (b) Calculated based on out-of-sample layers.}
	\label{fig:pair:hamming}
\end{figure}
The groups in Figure~\ref{fig:fao:wcloud} make intuitive sense for the most part, with some of the groups roughly corresponding to geographical regions. The analysis, however, reveals many unintuitive connections, with seemingly unrelated countries grouped together. Examples include: Macao and  Rwanda, Iraq and Guinea, Guam and Grenada, Fiji and Uganda, Bahrain and Mauritius, and so on. 

To verify these connections and the overall quality of clustering, we consider the following quantitative evaluation metric: For each pair of nodes, we compute the normalized Hamming distance for their corresponding rows in the adjacency matrices, and then take the average over layers. That is, we consider the following distance
\begin{align*}
	d(i,j) = \frac1T \sum_{t=1}^T \frac{1}{n_t} H(A_{ti*},A_{tj*}),
\end{align*}
with $H(\cdot, \cdot)$ denoting the Hamming distance. We refer to $d(i,j)$ as the Average Normalized Hamming (ANH) distance and note that it measures how dissimilar the connection patterns of two nodes are across layers. We expect nodes that are grouped together to have a lower value of $d(i,j)$ relative to a randomly selected pair. 

Figure~\ref{fig:pair:hamming}(a) shows plots of the distributions of $d(i,j)$ when the pairs are selected randomly within the groups versus the case were they are selected completely at random. The figure shows clearly that the ANH is much lower on average for within-group pairs relative to random pairs. This confirms that  HSBM  has uncovered groups based on trading patterns. 
The  median for ``Within'' and ``Random'' distributions are 0.027 and 0.21, respectively. The unintuitive pairs mentioned earlier have ANH much lower than the average for random pairs. For example, the ANH for Macao--Rwanda, Iraq--Guinea and Guam--Grenada,  Fiji--Uganda and Bahrain--Mauritius pairs are 0.074, 0.019, 0.0088, 0.14 and 0.14 respectively. This shows that these seemingly unrelated countries have similar trading pattern across multiple food product categories.

 In Figure~\ref{fig:pair:hamming}(a) the ANH is computed in-sample, that is, using the same 20 most dense layers used for fitting the HSBM.  Figure~\ref{fig:pair:hamming}(b) shows the corresponding plot when ANH is computed out-of-sample, using the remaining 344 layers. These layers are generally sparser, as reflected in the absolute value of ANH, but the plot shows a similar separation among the two distributions.

\section{Discussion}\label{sec:discuss}

In this work, we proposed a novel Bayesian  model for  community detection in multiplex networks by adopting the well-known HDP as a prior distribution for community assignments. Under the random partition prior,  a block model is assumed. This model facilitates flexible modeling of community structure as well as  link probabilities with its ability of incorporating potential  dependency and borrowing strength among networks  from different layers. For posterior inference of HSBM, we develop an efficient slicer sampler. The principles behind the slice  sampler can be applied to developing  sampling algorithms for many other models.  Future work will  be focused on developing models for community detection in networks with covariates, and  for inference of network-valued objects. 

\section*{Acknowledgement}

The contribution of L. Lin was funded by NSF grants DMS 1654579,  DMS 2113642 and a DARPA grant N66001-17-1-4041. A. Amini was partially supported by the NSF grant DMS-1945667.

\bibliographystyle{ba}
\bibliography{ref-LL-AA}

\end{document}